\documentclass[aps,prc,onecolumn,superscriptaddress,showpacs,floatfix]{revtex4}
\usepackage{graphicx,color}
\usepackage{amsmath}
\usepackage{amssymb}
\usepackage{bm}
\usepackage{amsfonts}%
\newcommand{\beq}{\begin{equation}}
\newcommand{\eeq}{\end{equation}}
\newcommand{\bea}{\begin{eqnarray}}
\newcommand{\eea}{\end{eqnarray}}
\newlength{\figwidth}

\begin{document}
\setlength{\figwidth}{0.98\columnwidth}

\title{The neutron within the deuteron as a surrogate for neutron-induced reactions}
\author{C. A. Bertulani}
\affiliation{Department of Physics and Astronomy, Texas A\&M University-Commerce, Commerce, TX, USA}
\affiliation{Department of Physics, Texas A\&M University, College Station, TX, USA}
\author{L. F. Canto}
\affiliation{Instituto de F\'{\i}sica, Universidade Federal Fluminense, Av. Litoranea
s/n, Gragoat\'{a}, Niter\'{o}i, R.J., 24210-340, Brazil and Instituto de F\'{\i}sica, Universidade Federal do Rio de Janeiro, CP 68528, Rio de
Janeiro, Brazil}
\author{M. S. Hussein}
\affiliation{Instituto de Estudos Avan\c{c}ados, Universidade de S\~{a}o Paulo C. P.
72012, 05508-970 S\~{a}o Paulo-SP, Brazil,\\
 Instituto de F\'{\i}sica,
Universidade de S\~{a}o Paulo, C. P. 66318, 05314-970 S\~{a}o Paulo,
Brazil,\\
Departamento de F\'{i}sica, Instituto Tecnol\'{o}gico de Aeron\'{a}utica, CTA, S\~{a}o Jos\'{e} dos Campos, S.P., Brazil}
\author{Shubhchintak}
\affiliation{Department of Physics and Astronomy, Texas A\&M University-Commerce, Commerce, TX, USA}
\author{T. V. Nhan Hao}
\affiliation{Department of Physics and Astronomy, Texas A\&M University-Commerce, Commerce, TX, USA}

\begin{abstract}
We propose the use of neutron poisons in reactions induced by radioactive beams as a test of theoretical models aiming to relate neutron capture in nuclei with neutron surrogate reactions such as (d,p) reactions. We exploit the approximations necessary to obtain a direct relation between the two reactions; surrogate vs. neutron capture. We also show how this is intimately related to the momentum distribution of the neutron within the deuteron. The models we use are based on the theory of inclusive breakup reactions commonly employed in the treatment of incomplete fusion and surrogate method. Such theories were developed in the 80's by Ichimura, Autern and Vincent [Phys. Rev. C 32, 431 (1985)],  Udagawa and Tamura [Phys. Rev. C 24, 1348 (1981)] and Hussein and McVoy [Nucl. Phys. A 445, 124 (1985)]. We use these theories to derive an expression for the proton yield in the reaction A(d,p)B. The capture reaction $n + A \rightarrow B$ is then extracted using reasonable approximations. By recalling an old method proposed by Serber [Phys. Rev. 80, 1098 (1950); Nature 166, 709 (1950)] we explain how the momentum distribution of neutrons within the deuteron will depend on the short-range dependence of the nucleon-nucleon force. The relevance of our work to nucleosynthesis in the rapid neutron capture process is emphasized.
\end{abstract}

\maketitle

\section{Introduction}

Neutron capture reactions are of both fundamental and practical importance \cite{BBF57}. Thermal neutrons are captured by a variety of targets, both stable and radioactive with a great variation in the value of the cross section. Most of these reactions cannot be determined directly due to the experimental difficulty of performing reactions with free neutrons. One often resorts to indirect methods which have proven to be of immense benefit for nuclear astrophysics and for nuclear science in general \cite{BG10,BK16}. One of such indirect methods uses transfer reactions to induce reactions in which the neutron is a participant. These are often called ``{\it surrogate reactions}" in which the neutron is carried within a ``trojan" projectile and brought to react with the target \cite{CB70}.  For example, in (d,p) reactions one hopes  to infer neutron-induced cross sections on several nuclear targets by measuring the proton in the final stage \cite{ED06,Fors07}. The surrogate, or trojan, method is even more important when studying neutron-induced reactions on  short-lived nuclei because it is extremely difficult to handle radioactive samples. But despite numerous efforts to relate (d,p) and other surrogate reactions with those induced by free neutrons, this research field is still somewhat murky, with some lingering problems still remaining. Perhaps the most difficult one is due to the fact that the neutron within the deuteron will populate different angular momenta in the target as compared to those populated by free neutrons. Due to the sensitivity of the cross sections on the angular momentum, this is a critical issue \cite{DJ15}. One believes that when the compound nucleus reaction induced by a surrogate can be well explained using the  Weisskof-Ewing method, and not necessarily the Hauser-Feshbach method, then the reaction is closely related to that induced by a free neutron \cite{BJ11}.

Apart from the angular momentum matching issue,  perhaps the simplest and most difficult problem in (d,p) and other surrogate reactions is to quantify theoretically how parts of the neutron wavefunction within the deuteron are of importance in the reaction. This question might lack clarity because one does not really observe wavefunctions. Theorists think about wavefunctions while experimentalists only can test observables.  Of course, if one had a perfect reaction theory, such a question might not even make sense, as one had all the quantum aspects of the problem theoretically under control for the please of theorists and experimentalists alike. But the fact is that there is no accurate theory to quantify how well surrogate reactions work and what details of the neutron wavefunction within the deuteron emerge in the reaction process.   The sample of cases which have been studied so far does not allow to discern what are the best theories for surrogate reactions, as very phenomenological models are used including optical potentials containing hidden information that are not transportable from one reaction problem such as elastic scattering to  another such as nuclear excitation some particular channel. Despite such challenges, there are very few experimental alternatives to infer neutron-induced reactions of relevance for nuclear science. Surrogate reactions are indeed the most promising tool, and maybe the only one, to replace reactions with free neutrons, and have had well documented successes \cite{Burk06,TH07,Bas09,Alm09,Nay14}.

Theorists can steadily increase the complexity model for surrogate reactions, such as using sophisticated three-body scattering models, e.g., the Alt-Sandhas-Grassberger (AGS) method to tackle the three-body nature of the reaction mechanism in (d,p) reactions\cite{GrassSand67,AGS67} . To it one can also include more complexities due to excitation of specific nuclear states \cite{MES12}. But  very basic questions about the connection of the two distinct neutron wavefunctions, namely, the bound-state in the deuteron and the free neutron wavefunction, have not been explored to content in the literature. In this article we plan to tackle exactly this part of the  problem: what are the necessary qualifications of the neutron wavefunction within a surrogate nucleus to perform the same tasks as the free neutron? To simplify the problem even further, we will not dwell on what exactly the free neutron will do with the nuclei. We just assume it does whatever it does. To be a bit more specific we will assume that there is a cross section number coming from measurements of (free) neutron reactions.  Then we ask what are the conditions needed to relate (d,p) reactions to neutron reaction cross sections.  Another relevant issue is to find out how to relate the thermal neutron energies in the laboratory to those of the neutron within the deuteron. Most of the existing radioactive beam facilities provide nuclei at several tens, or hundreds, of MeV per nucleon which can be used to react with deuteron targets. This translates, within a factor of 2, to a similar nominal energy for the projectile deuteron in inverse kinematics. The neutron within the deuteron, captured at such energies, will be useless for most cases of interest for neutron capture in stars or other applications of nuclear science. Without a firm understanding of the parts of the neutron within the deuteron which are of significance for surrogate reactions deems such reactions as basically useless for practical applications. 

To  gain insight on such problems we do not need to use the most sophisticated theories, but those who already include most of the physics ingredients. The best experimental scenario would involve the largest possible neutron reaction cross sections, as it would in principle also lead to the largest surrogate cross sections. This is the case of neutron poisons, for which the reactions cross sections are huge, sometimes as large as 3 million barns for thermal neutron capture on $^{135}$Xe. It is well under experimental reach to produce secondary beams of $^{135}$Xe nuclei in fragmentation reactions and guide them to a deuteron target.  Because of its huge cross section for free neutrons, we also expect a large (d,p) cross section on the same nucleus. With large number of events and good experimental statistics this reaction, and other reactions with neutron poisons, would therefore constitute a perfect benchmarking for tests of surrogate reactions. Such a test is crucial for determining if much smaller cross sections  involving radioactive beams can be used in conjunction with this technique to determine neutron capture reactions of interest for nuclear astrophysics, such as those of relevance for the r-process.

This paper is organized as follows. In Section II we first give a summary of the inclusive breakup theory as developed in Refs. \cite{UT1981, UT1984, IAV1985, HM1985, Austern1987}. We use the simplest of these, developed in Ref. \cite{HM1985}, which does not require  the solution of an equation beyond the usual homogeneous optical Schr\"{o}dinger equation, as do the prior-form theory of Udagawa and Tamura, \cite{UT1981} and the post-form theory of Ichimura, Austern and Vincent \cite{IAV1985}, to the (d,p) reactions on radioactive targets. The Hussein-McVoy theory has already been successfully applied to analyze breakup data in Refs. \cite{HFM1990,Ichimura1990,Moro2015,HM1989, BM1992, Canto1998}. Our sole goal in this Section is to show under what circumstances the (d,p) reaction can be found proportional to the neutron reaction cross section. In the case of very short-lived exotic neutron-rich nuclei, such as the one-neutron halo, $^{11}$Be (half-life = $13.76 \pm 0.07$ s), and the two-neutron halo, $^{22}$C (half-life = $9\pm 3 $ ms), an inverse reaction must be considered, i.e., d(Be,X)p. This would be the same situation if one uses neutron poisons as secondary beams. Concomitantly, in the Appendix we use the Serber\&Butler model to determine the approximations needed to prove that the knowledge of the momentum distribution of the neutrons within the deuteron is fundamental for the calculation of surrogate neutron cross sections. Based on these results, we introduce a practical method to estimate the (d,p) cross sections with the neutron capture at thermal energies. These are the only energies of interest for astrophysics. The other, much higher, energies are to be considered as ``{\it garbage}". In Section IV we discuss the relevance of neutron poisons for nuclear science and in Section IV we apply our formalism to the $^{135}$Xe(d,p)$^{156}$Xe and other surrogate reactions. We propose to use them as benchmarks to test the effect of hybridity of the process. We present our extracted capture cross sections for the radioactive nuclei at thermal energies, and at energies encountered in the r-process, $E_n \lesssim 20-30$ keV. Finally, in Sections V and VI we present a brief discussion and our concluding remarks.

\section{Factorization of inclusive breakup reactions}

\subsection{General theory}
In this Section we discuss the process in which a part of the projectile is captured by the target nucleus.  The remaining part of the projectile is considered to be a spectator and merely scatters off, such as in a (d,p) reaction (Although we envisage deuterons as targets in radioactive beam facilities, we term them as projectiles based on the idea of working with them in inverse kinematics.). The experimental observation involves a spectrum of this spectator fragment and our aim is to extract from the measured spectrum the capture cross section of the participant fragment. We denote the projectile by $a = b + x$, where $b$ is the observed spectator, while $x$ is the captured fragment. The target is denoted by $A$. The exact Hamiltonian of the three-body system, $ x + b + A$ is given by,
\begin{equation}
\Big[E - \big( K_a + K_x + K_A+ V_{xb} +V_{xA} + V_{bA} + h_A\big)\Big] \left|\Psi^{(+)}\Phi^{x}_{A}\phi_a\right> = 0 ,
\end{equation}
where $K$ is the kinetic energy operator, $V_{ij}$ is real interaction potential between the pair $ij$, and $h_A$ is the intrinsic Hamiltonian of the target nucleus $A$. The wave functions $\Psi^{(+)}, \Phi^{x}_{A}$  and $\phi_a$ are, respectively, the exact scattering wave function describing the relative motion of the $x + b + A$, the intrinsic wave functions of the target, and of the projectile. The first approximation we make is to neglect $K_A$ as very little recoil the target will suffer since in most cases we consider, $a/A \ll 1$. The second approximation is the spectator one, $V_{bA} \rightarrow U_{bA}$, where $U_{bA}$ is the complex optical potential of the $b$ fragment in the complex field of the target. The reaction to be considered is $ a + A \rightarrow b + X$, where $ a = b + x$ and the final nucleus is $X = x + A$. For (d,p) reactions, x = n (neutron) and b = p (proton). For simplicity, we will not make explicit mention of angular momenta quantum numbers.

The expression for the double differential cross section of $b$ being observed has been derived by several authors \cite{UT1981,UT1984,HM1985,IAV1985},
\begin{equation}
\frac{d^{2}\sigma}{d\Omega_{b}dE_{b}} = - \frac{2}{\hbar v_a} \rho_{b}(E_b) \left<\hat{\rho}_x\left|W_{x}\right|\hat{\rho}_x\right> \label{IC} ,
\end{equation}
where $\rho_{b}(E_b) = k_{b}\mu_{b}/[(2\pi)^3 \hbar^2] $ is the density of states of the outgoing proton (with $\mu_b$ being the reduced
mass of b + X, and $k_b = \sqrt{2\mu_{b}E_b/\hbar^2}$ being their relative wave number. Here $E_b = E_a - E_x$. The  fragment $x$ source function $\hat{\rho}_{x}({\bf  r}_x)$ is calculated using the post representation by Ichimura-Austern-Vincent (IAV) \cite{IAV1985}, $\hat{\rho}_{x, IAV}({\bf r}_x)$, the prior representation by Udagawa-Tamura (UT) \cite{UT1984},  $\hat{\rho}_{x, UT}({\bf r}_x)$ and  the Hussein-McVoy (HM) \cite{HM1985} one is the difference, $\hat{\rho}_{x, HM}({\bf r}_c)=\hat{\rho}_{x, IAV}({\bf  r}_x) - \hat{\rho}_{x, UT}({\bf r}_b)$, as was demonstrated in Refs. \cite{HFM1990, Ichimura1990}. The equations satisfied by these source functions are usually evaluated within the distorted wave Born approximation (DWBA). In Eq. (\ref{IC}), $W_x$ is the imaginary part of the $xA$ optical potential.

The post IAV source function equation is \cite{Moro1-2015, Moro2015}
\begin{equation}
 \Big[E_x - K_x - U_{xA}\Big]\hat{\rho}_{x, IAV}({\bf r}_x)= \left(\chi_{b}^{(-)}\right|
 \Big[V_{bx} + U_{bA} - U_{bX}\Big]  
 \left|\chi_{d}^{(+)}\phi_{d}\right> \approx \left(\chi_{b}^{(-)}\right|V_{bx} \left|\chi_{a}^{(+)}\phi_{a}\right>,
 \label{IAVeq}
 \end{equation}
 and the prior UT source function equation,
 \begin{equation}
\Big[E_x - K_x - U_{xA}\Big]\hat{\rho}_{x, UT}({\bf r}_x) = \left(\chi_{b}^{(-)}\right|\Big[U_{bA} + U_{xA} - U_{aA}\Big]\left|\chi_{a}^{(+)}\phi_{a}\right>.
\label{UTeq}
\end{equation}
The HM source function satisfies an equation which is just the difference between Eqs. \eqref{IAVeq} and \eqref{UTeq},
\begin{equation}
\Big[E_x - K_x - U_{xA}\Big]\hat{\rho}_{x, HM}({\bf r}_x) \approx \left(\chi_{b}^{(-)}\right|\Big[V_{bx} - \left(U_{nA} + U_{pA} - U_{dA}\right)\Big]\left|\chi_{a}^{(+)}\phi_{a}\right> .\label{HMeq}
\end{equation}
In the above $\phi_{a}(\textbf{r}_{bx})$ is the internal wave function of the projectile, $\chi_{a}(\textbf{r}_{dA})$ and $\chi_{b}(\textbf{r}_{bA})$  are the distorted waves of the incoming projectile, $a$, and the outgoing spectator fragment, $b$.
These wave functions are distorted by the complex optical potentials, $U_{aA}({\bf r}_{aA})$, and  $U_{bA}({\bf r}_{bA})$, respectively. The outgoing fragment, $b$,  is also distorted by the complex field of the final nucleus $X$, through the potential $U_{xX}({\bf r}_{xX})$.

The IAV cross section is then a sum of three terms,
\begin{equation}
\frac{d^2\sigma_{IAV}}{d\Omega_{b}dE_{b}}  = \frac{d^2\sigma_{UT}}{d\Omega_{b}dE_{b}} + \frac{d^2\sigma_{HM}}{d\Omega_{b}dE_{b}} + \frac{d^2\sigma_{Int}}{d\Omega_{b}dE_{b}} .
\end{equation}
The last term is the interference contribution. The physical interpretation of the the UT and HM contributions was analyzed by several authors. The UT cross section corresponds to the explicit two-step  process of elastic breakup of $a$ followed by the capture of fragment $x$ by the target. The HM cross section represents the capture with concomitant excitation of the target, and the inclusive direct excitation of the target by neutron. Clearly, the IAV cross section contains all these contributions and thus is the most inclusive. Recently, Ref. \cite{Moro2015} made a careful analysis of the inclusive cross section by calculation each of the contributions. At higher energies, the UT cross section is  small, and a reasonable approximation is to set the IAV cross section equal to the HM one. The advantage of the HM cross section is that the source function is just the non-orthogonality function $\hat{\rho}_{x, HM}({\bf r}_x) = \left(\chi_{b}^{(-)}\big|\chi_{a}^{(+)}\phi_{a}\right> ({\bf r}_x)$. There is no need to solve the inhomogeneous equation Eq. (\ref{HMeq}).

If the inhomogeneous terms in Eqs. (\ref{IAVeq}, \ref{UTeq}, \ref{HMeq}),  were to be ignored, the $x$ fragment source function becomes just the $x$ distorted wave with the $xA$ optical potential. The cross section, Eq. (\ref{IC}) then becomes the product of the density of states available for the spectator fragment, $b$,  $\rho_{b}(E_b)$ times the total reaction cross section of $x$ with the target. No reference to the incoming deuteron is maintained except the trivial energy relation $E_n = E_d - E_p$.

For the purpose of this paper, namely studying neutron capture through the (d,p) reaction, it is important to assess the difference between the free $xA$ total reaction cross section and the one extracted from the (d,p) reaction. We develop this in the following.

\subsection{Deuteron induced neutron capture}
Here we will show how for the case of the deuteron, the cross section for (d,p) reaction can be related to that for neutron-induced reactions. We use the Hussein-McVoy \cite{HM1985, HM1989, BM1992} expression for the cross section, as it is most appropriate for our purpose, being based on DWBA approximation to the full three-body wave function representing the entrance channel \cite{Canto1998}, and with a source function which does not require the solution of an inhomogeneuous equation, such as Eq. (\ref{HMeq}),
\begin{equation}
\hat{\rho}_{n}({\bf r}_n) = \left(\chi^{(-)}_{p}({\bf r}_p)\Big|\chi^{(+)}_{d}({\bf r}_n, {\bf r}_p) \phi_{d}({\bf r}_{p} - {\bf r}_{n})\right> ,\label{source}
\end{equation}
where $\chi$ is the optical model wave function (distorted wave), and $\phi_d$ is the intrinsic wave function of the deuteron.

The above expression is obtained in the post representation where the interaction in the final state is $V_{n,p}$. Note that the energy available for the $nA$ system is $E_i + B_d -E_p$, where $E_i$ is the incident energy in the center of mass system of $d + A$, $B_d$, is the binding energy of the projectile considered as a $n + p$ system, and $E_p$ is the variable energy of the observed proton. The total reaction cross section of the system $n + A$ system in a $n + A \rightarrow B$ reaction is
\begin{equation}
\sigma_{R}^{nA} = \frac{k_n}{E_n} \left<\chi_{n}^{(+)}({\bf k}_n)\left|W_{nA}(E_n)\right|\chi_{n}^{(+)}({\bf k}_n)\right>. \label{sigmax}
\end{equation}
Next we prove that a explicit relation between Eq. (\ref{IC}) and Eq. (\ref{sigmax}) exists.

To exhibit this relation we have to analyze the source function $\Big(\chi^{(-)}_{p}({\bf  r}_p)\Big| \chi^{(+)}_{d}({\bf  r}_n, {\bf r}_p) \phi_{d}({\bf r}_{p} -{\bf r}_{n})\Big>$. If we use the approximation 
\begin{equation}
\left| \chi_{d}^{(+)}({\bf r}_p,{\bf r}_n)\right> = \left|\chi_{p}^{(+)}({\bf r}_{p})\right> \left|\chi_{n}^{(+)}({\bf r}_{n})\right>,
\end{equation}
the source function becomes,
\begin{equation}
\Big(\chi^{(-)}_{p}({\bf r}_p)\Big|\chi^{(+)}_{d}({\bf  r}_n, {\bf r}_p) \phi_{d}({\bf r}_{p} -{\bf r}_{n})\Big> = \int d{\bf r}_{p} S_{p}({\bf r}_p) \phi_{d}({\bf r}_{p}-{\bf r}_{n})\chi^{(+)}_{n}({\bf r}_{n}),
\end{equation}
where $S_{p}({\bf r}_p)\equiv S_{{\bf k}^{\prime}_{p},{\bf k}_{p}}({\bf r}_{p})$ is the ${\bf r}_p$ integrand in the integral that defines the elastic $S$-matrix element of the spectator proton,
\begin{equation}
S_{{\bf k}^{\prime}_{p},{\bf k}_{p}} = \int d{\bf r}_{p} S_{{\bf k}^{\prime}_{p},{\bf k}_{p}}({\bf r}_{p}).
\end{equation}
The inclusive non-elastic proton spectrum in the A(d,p)B reaction is then,
\begin{equation}
\frac{d^{2}\sigma}{d\Omega_{p}dE_{p}}= \frac{k_{d}}{E_{d}}\rho_{p}(E_p)\int d{\bf r}_{p} d{\bf r}_{p}d{\bf r}^{\prime}_{p}\phi_{a}^{\star}({\bf r}_p - {\bf r}_{n})\phi_{d}({\bf r}^{\prime}_p - {\bf r}_{n}) S_{p}^{\star}({\bf r}^{\prime}_{p})S_{p}({\bf r}_{p}) |\chi^{(+)}_{n}({\bf r}_{n})|^{2}W_{{\bf r}_n}({\bf r}_{n}).
\end{equation}

We now introduce the modified  proton elastic $S$-matrix,
\begin{equation}
\hat{S}_{p}({\bf r}_{n}) \equiv \int d{\bf r}_{p} \phi_{d}({\bf r}_{p} - {\bf r}_{n}) S_{p}({\bf r}_{p}).
\end{equation}
Then the inclusive non-elastic (capture) cross section is, for the protons,
\begin{equation}
\frac{d^{2}\sigma}{d\Omega_{p}dE_{p}} =  \frac{k_{d}}{E_{d}}\rho_{p}(E_p) \int d{\bf r}_{n} |\hat{S}_{p}({\bf r}_{n})|^{2}|\chi^{(+)}_{n}({\bf r}_{n})|^{2} W_{n}({\bf r}_{n})
= \frac{k_{d}E_{n}}{k_{n}E_{d}} \rho_{p}(E_p) \hat{\sigma}^{nA}_{R}, \label{sigfact}
\end{equation}
where the modified total reaction cross section, $\hat\sigma^{nA}_{R}$, is
\begin{equation}
\hat{\sigma}^{nA}_{R} = \frac{k_n}{E_n} \int d{\bf r}_{n} |\hat{S}_{p}({\bf r}_{n})|^{2}|\chi^{(+)}_{n}({\bf r}_{n})|^{2} W_{n}({\bf r}_{n}),\label{sigmaxM}
\end{equation}
to be compared to Eq.(\ref{sigmax}),
\begin{equation}
\sigma^{nA}_{R} = \frac{k_n}{E_n}\int d{\bf r}_{n} |\chi^{(+)}_{n}({\bf r}_n)|^{2} W_{nA}({\bf r}_{n}). \label{sigmax1}
\end{equation}

Eq. (\ref{sigmaxM}) is quite similar to Eq. (\ref{sigmax1}), for the total reaction cross section of the neutron. The major difference is the presence of the $\phi_{d}$-modified elastic S-matrix of the proton. Two factors enter in the difference between the two equations. The internal wave function of the deuteron, and the elastic proton $S$-matrix. Of course there is also the constraint on the energy of the neutron arising from $E_n = E_d +B_d -E_p$.   This proves our assertion that the (d,p) cross section can be related to the neutron-nucleus reaction cross section, under certain approximations.  It also becomes clear how the ideal case of a direct proportionality of surrogate and free neutron capture is spoiled by the presence of distortion effects within the matrix elements. In the Appendix we show how to link the cross sections for surrogate reactions with the momentum distributions of the neutron within the deuteron.

 As demonstrated in this Section, it is also clear that the geometric features of the reaction, manifested through the reaction S-matrices, modify the simple picture presented in the Appendix. The lesson we had learned along the way is that under certain circumstances the (d,p) cross sections  can be proven proportional to the neutron-capture cross sections and at the same time it is also dependent on a geometrically modified momentum dependence of the deuteron wavefunction.  In the next Section we will explore both of these features to assess the important question on how much the (d,p) reaction contains information on neutron capture cross sections. We build our case on the results obtained in this Section as well as on that presented in the Appendix.    

\subsection{Neutron surrogate factor and energy matching}

In this Section we discuss a practical and transparent method to estimate the (d,p) cross section in terms of the neutron-induced reaction cross section. To probe the very low energy neutron capture through the (d,p) reaction, one has to look at the high energy tail of the neutron energy spectrum within the deuteron. At the most common high deuteron energies obtained in many present nuclear physics facilities, one may venture and use the eikonal approximation for $S_{p}({\bf r}_{p})=\exp[i\chi(b_p)]$, where
\begin{equation}
\chi_{p}({\bf b}_{p}) = i\frac{k_p}{2E_p}\int_{-\infty}^{\infty} dz \ U_{p}(z, {\bf b}_p), \label{chibp}
\end{equation}
and to it one also needs to add the Coulomb phase \cite{ABH92}. In the above ${\bf b}_p$ is the two-dimensional proton impact parameter vector. For deuterons at 50 MeV and above, one can also resort to the same eikonal approximation for the neutron. In this case, very transparent expressions emerge which help us to quantify the relation between (d,p) and neutron-induced reactions. We follow some of the ideas laid down in the previous Section, the Appendix, and in Ref. \cite{HM85}. Note that the eikonal approximation is not justifiable for the lower part of the deuteron energies we will discuss in this article. Its use makes the expressions very easy to understand, allowing for a probabilistic  interpretation of the reaction cross section. A more justifiable procedure would be a partial wave expansion of the scattering wavefunctions. But the expressions one obtains in that case lack clarity for what we want to explore. 
  
\begin{figure}
\begin{center}
\includegraphics[scale=0.35]{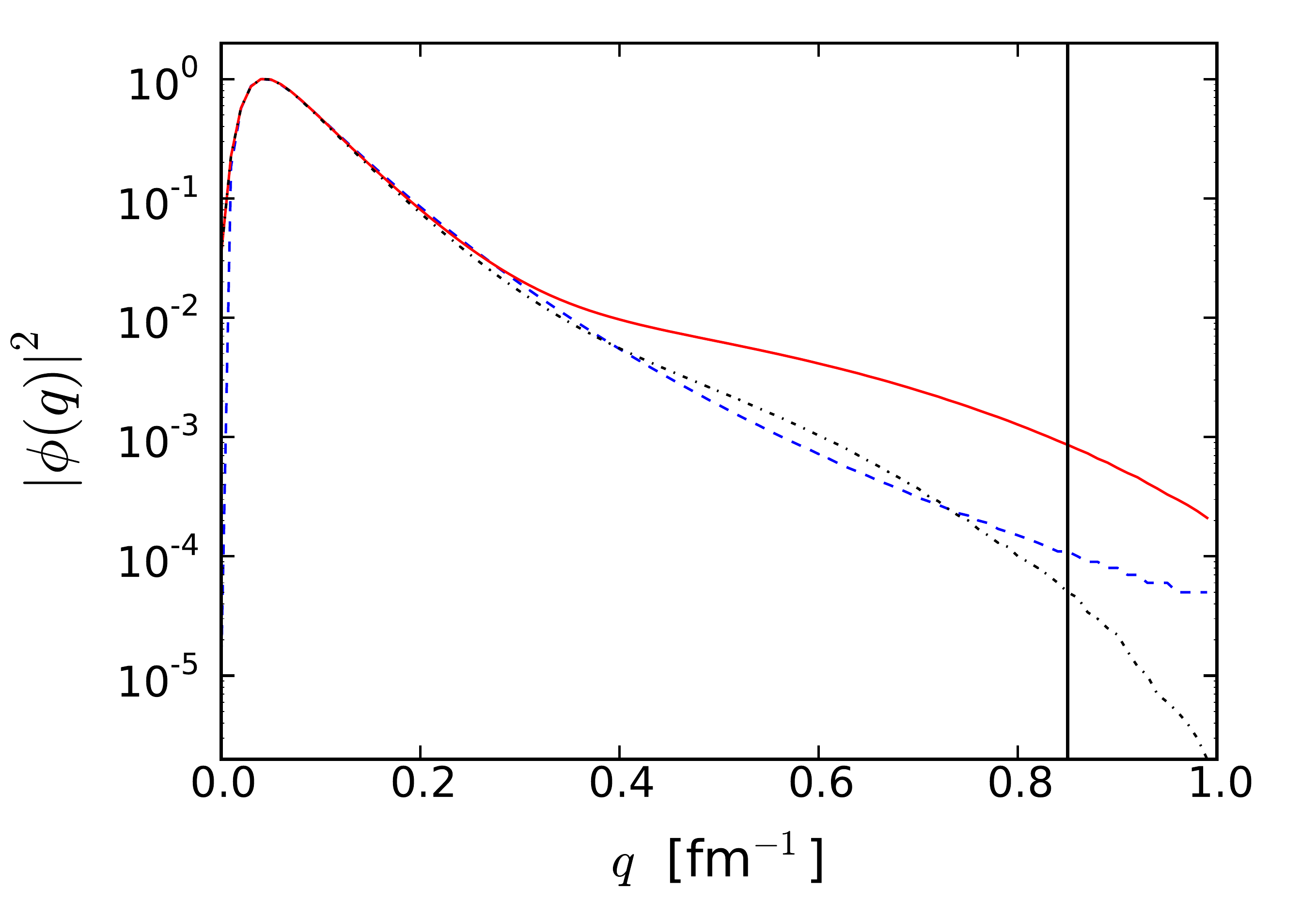}
\caption{Three results for the deuteron nucleon momentum distribution. The Hulthen momentum distribution (dashed line), and the corresponding distributions from the Bonn (solid line) and from the Paris (dotted line) potentials. For a deuteron projectile with  30 MeV, only the tail of the momentum distributions above the vertical line yield thermal neutrons in the laboratory.} \label{f1}
\end{center}
\end{figure} 

In the (d,p) reactions at moderately high energies ($\gtrsim 50$ MeV), one can use the eikonal S-matrices to determine the neutron removal, $1-|S_n({\bf b}_n)|^2$, and the survival probability for the proton, $|S_p({\bf b}_p)|^2$, where $b_n$ and $b_p$ are the neutron and proton coordinates transverse to the beam direction, usually interpreted as their impact parameters in the classical limit.  One also has to account for the probability of finding the neutron within a distance ${\bf r} = {\bf r}_n-{\bf r}_p$ from the proton, which is given by $|\phi_d({\bf r}_n-{\bf r}_p)|^2$. In this formalism, the cross section for the proton survival with the neutron being absorbed by target nuclei (or by the projectiles in reactions with secondary beams) is given by
\begin{equation}
\sigma_{-n} = \int d^2b_n d^2b_p dz_p \left|S_p({\bf b}_p)\right|^2    \left( 1 -\left|S_{n}({\bf b}_{n})\right|^2 \right) \left|\phi_d({\bf r}_n-{\bf r}_p)\right|^2. \label{signmn}
\end{equation}
These integrals are entangled through the deuteron wavefunction, and the coordinate definition ${\bf r}_n-{\bf r}_p \equiv ({\bf b}_n - {\bf b}_p, z_n - z_p)$. The relation to the neutron absorption cross section is easily seen as a consequence of relaxation of the ``entanglement" of the neutron. That is, if the integration over the neutron variable $b_n$ would be factored out, the neutron would act as a free particle. To prove this assertion, let us assumes that $\left|\phi_d({\bf r}_n-{\bf r}_p)\right|^2 = 1/V$, where $V$ is an irrelevant normalization volume for the p-n free wave wavefunction. One then gets
\begin{equation}
\sigma_{-n}^{free} = \int d^2b_n \left( 1 -\left|S_{n}({\bf b}_{n})\right|^2 \right) \left[ {1\over V} \int d^2b_p dz_p \left|S_p({\bf b})\right|^2 \right]   = \sigma^{nA}_G \left[ {1\over V} \int d^2b_p dz \left|S_p({\bf b})\right|^2 \right]    , \label{signfr}
\end{equation}
where $\sigma^{nA}_G$ is the geometric neutron-nucleus cross section. It is nearly the same as the cross sectional area of the matter density in the nucleus. This equation has a similar form as Eq. \eqref{sigfact}, but with one caveat: $\sigma^{nA}_G$ here is by no means equal to $\sigma^{nA}_{R}$ in Eq. \eqref{sigmax1}, the ``de facto" neutron-nucleus cross section. However, Eqs. \eqref{signmn} is very useful in that it determines how much of the neutron within the deuteron is free to react with the nucleus $A$. 

The above discussion shows that one can estimate the (d,p) cross section by defining a {\it neutron surrogate factor}, $P_{n/d}$, as
\begin{equation}
P_{n/d} (E_d)= {1\over \sigma^{nA}_{G}}\int d^2b_n d^2b_p dz_p \left|S_p({\bf b}_p)\right|^2    \left( 1 -\left|S_{n}({\bf b}_{n})\right|^2 \right) \left|\phi_d({\bf r}_n-{\bf r}_p)\right|^2.   \label{Pnd}
\end{equation} 
The neutron surrogate factor incorporates the fact that the collision chops off a chunk the deuteron wavefunction to make it available for the neutron capture. This is encoded in the geometrical factors $ \left|S_p({\bf b}_p)\right|^2    \left( 1 -\left|S_{n}({\bf b}_{n})\right|^2 \right)$. It involves the tail of the matter distribution associated with the deuteron wavefunction in grazing collisions. The rest of the deuteron wavefunction is lost forever in the ``{\it reaction trash}", namely, it becomes useless for neutron capture purposes. In contrast to the direct proportionality to the neutron momentum distribution in the surrogate projectile, the geometrical factors $ \left|S_p\right|^2    \left( 1 -\left|S_{n}\right|^2 \right)$ also constrain most of the available momenta of the neutron within the deuteron to those small momenta matching the low energy requirement to induce large thermal cross sections.

Comparing to Eqs.  (\ref{sigfact}-\ref{sigmax1}) we see that the factor $  1 -\left|S_{n}({\bf b}_{n})\right|^2 $ corresponds to $|\chi^{(+)}_{n}({\bf r}_n)|^{2} W_{nA}({\bf r}_{n})$. The free neutron distorted wave is modified (absorbed) by the neutron-nucleus optical potential $W_{nA}({\bf r}_{n})$ at small distances in a similar way as the $ 1 -\left|S_{n}({\bf b}_{n})\right|^2 $ term does for Eq. \eqref{signfr} in high-energy collisions. It is also evident that the difference between the reaction with a free neutron with that involving a neutron within the deuteron is manifested through the proton survival requirement in the factor $\left|S_p\right|^2$. Indeed, as we showed in Eqs. \eqref{sigmaxM} and \eqref{sigmax1}, it is precisely this difference which sends to the reaction trash most of the neutron wavefunction within the deuteron. 

In order to be useful for estimates of neutron capture at thermal energies,one needs to estimate how much of the deuteron wavefunction contains neutrons with thermal energies in the laboratory frame of reference. The momentum distribution, or Fermi motion, of the neutron within the deuteron is here denoted by $|\phi({\bf q})|^2$.   From this distribution one has to  select the momenta ${\bf q}$ corresponding to laboratory energies $E_n < E_n^{th}$. Notice that we use the term ``thermal energy" in a loose way: for reactor physics, a thermal neutron has about 0.025 eV, whereas for nuclear astrophysics their energy is in the range of a few tens of keV. These are the most probable energies according to  the Maxwell-Boltzmann distribution for each of the relevant temperatures. In any of these situations, the thermal neutron will have energies much smaller than the projectile deuterons.  

Let us initially consider s-waves for the deuteron. In Figure \ref{f1} we show three results for the deuteron nucleon momentum distribution obtained with (a) the Hulthen \cite{Hul57} momentum distribution (dashed line), and the corresponding distributions from the Bonn \cite{MHE87} (solid line) and from the Paris \cite{Lac81} (dotted line) potentials. For a deuteron projectile with  50 MeV, only the tail of the momentum distributions above the vertical line yield thermal neutrons in the laboratory. For such values the relative number of neutrons available for capture at thermal energies in the laboratory decreases by a factor $10^{-5}-10^{-3}$ depending on the neutron wavefunction used. At such high momenta, the distribution is also very sensitive to the short distance features of the deuteron wave function, including the effects of short-range part of the  interaction.  Therefore, in contrast to what has been discussed in the literature so far, surrogate reactions at radioactive beam facilities  are complicated by the lack of a good understanding of short-range correlations in nuclei.  Indeed, the topic of short-range correlations in nuclei has attracted a large interest in recent years, due to its connections to problems as far-reaching as the equation of state of nuclear matter in neutron stars \cite{Eli12,Hen14}. Its relation with surrogate reactions has not been explored so far.

\begin{figure}
\begin{center}
\includegraphics[scale=0.35]{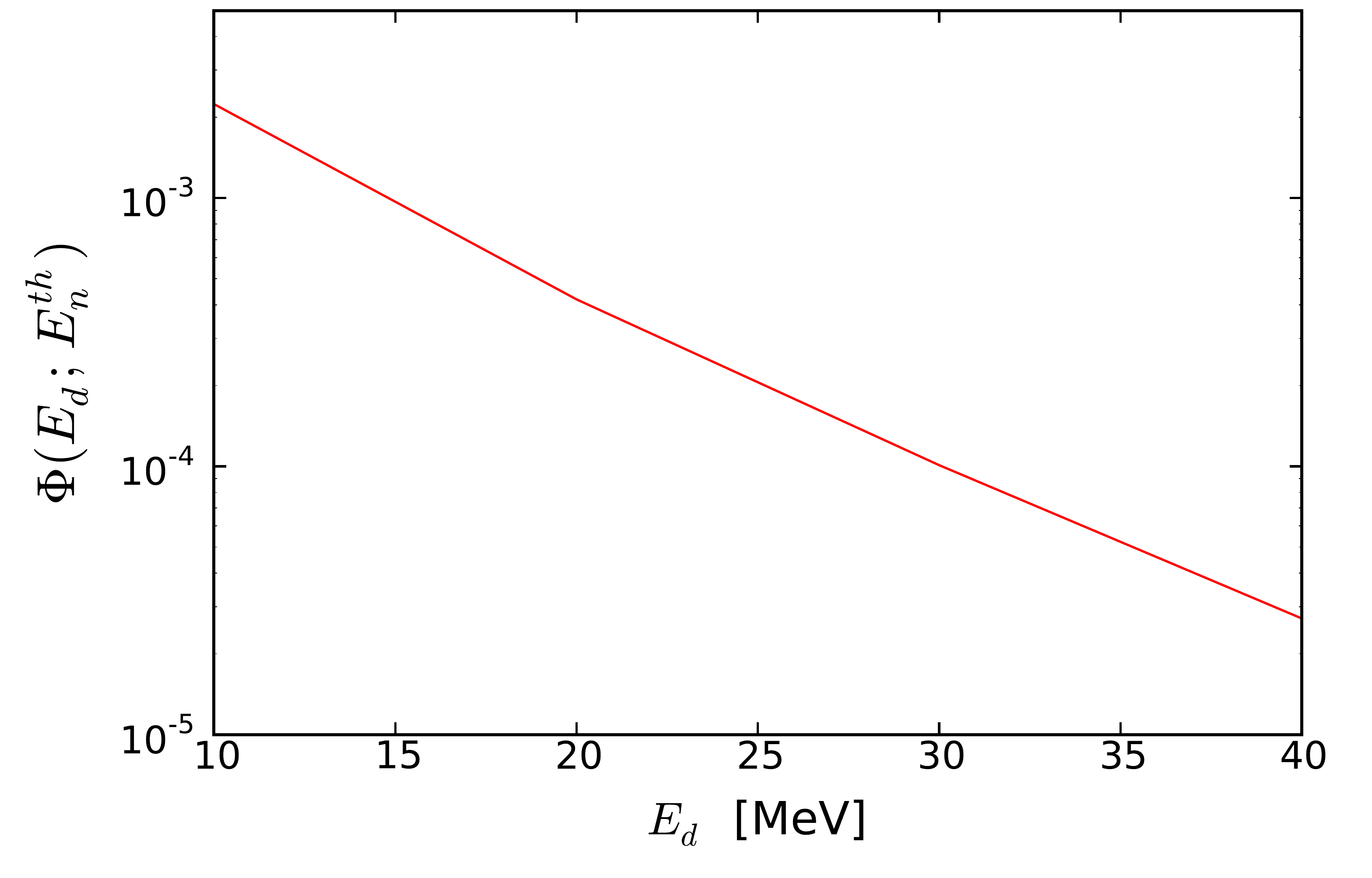}
\caption{Momentum matching function (probability) for the capture of thermal neutrons by the target for a deuteron projectile with energy $E_d$.} \label{f2}
\end{center}
\end{figure}

We define the ``{\it momentum matching function}"
\beq
\Phi(E_d;E_n^{th})={1\over 6}
{\int_{E_d}^\infty |\phi({\bf q})|^2 d^3 q \over \int_{0}^\infty |\phi({\bf q})|^2 d^3 q}, \label{match}
\eeq
as a measure of the thermal neutrons seen in the laboratory for deuterons with energy $E_d$.  The factor 1/6 accounts for 1/3 of the neutrons moving along the beam direction, and only 1/2 of them with momenta greater than zero in the laboratory. Notice that, since $E_n^{th} \ll E_d$, the lower limit of the integral in the numerator can be set to $E_d$ instead of $E_d \pm E_n^{th}$ without changing appreciably the results. The function \eqref{match} effectively measures the probability that very low laboratory energy neutrons are available in surrogate reactions. We will use the Hulthen \cite{Hul57} to calculate the matching function, as displayed in Figure \ref{f2}. We observe that the momentum matching function (probability) for the capture of thermal neutrons by the target for a deuteron projectile with energy $E_d$ decreases steadily with the deuteron energy, as expected.

In terms of the neutron surrogate factor and the momentum matching probability, the (d,p) cross section becomes a fraction of the n-A reaction cross section, and is determined by the product
\begin{equation}
\sigma_{(d,p)}^{s(d)}(E_n)=P_{n/d}^{s(d)} (E_d) \cdot \Phi(E_d;E_n^{th}) \cdot \sigma^{nA}_{R}(E_n), 
\end{equation} 
where $P_{n/d}^{s(d)}(E_d)$ is the neutron surrogate factor for the $s$ ($d$) parts of the deuteron wavefunction, calculated for incident deuteron energy $E_d$. We will use this equation with the experimental values of $\sigma^{nA}_{R}(E_n)$ to calculate (d,p) reactions for which the neutron is captured at energies $E_n$. We agin emphasize that, while this method is justifiable for high energy deuterons, at a few tens of MeV and higher, it is not thought to be appropriate at lower energies. However, it serves as a guidance for a quite complicated problem, which so far has not found a robust and accurate theoretical treatment in the literature. 

The above discussion highlights the key problems in using deuteron as a surrogate for neutron-induced reactions at thermal energies. Indeed, since the matching function decreases with increasing deuteron energy, one might expect that the best scenario would be to carry out experiments at low deuteron energies. However, due to the Coulomb repulsion the neutron capture will also decrease due to a smaller overlap of the deuteron wavefunction with the target. We will prove this assertion in a later Section.

\section{Neutron poisons} 
In the case of the stable $^{10}$B and $^{157}$Gd, the neutron induced cross sections are, respectively, $3.80 \times 10^3$ barns and $2.54 \times 10^5$ barns. Such large values, combined with their chemical properties, make them quite valuable in medical applications, such as the Boron Neutron Capture Therapy (BNCT) \cite{YF13} and the Gadolinium Neutron Capture Therapy (GNCT) \cite{Nov13}. Other cases of large thermal neutron capture cross sections are $^{153}$Cd, $2 \times 10^{4}$ barns, and $^{135}$Xe, $3 \times 10^{6}$ barns. The cadmium isotope $^{113}$Cd is a reactor ``poison"  and is commonly used as a neutron absorber-moderator in reactors and in other applications. In the case of the radioactive nucleus $^{135}$Xe (whose lifetime is about 10 hours), the very large capture cross section  reaches atomic values. Nuclei in the vicinity of these isotopes exhibit  much smaller cross sections. To give an idea of the values of the thermal neutron capture cross sections on selected nuclei, we reproduce an empirical list adapted from Ref. \cite{Mughab2003} in Table 1.  Data for Xe isotopes were taken from Refs. \cite{Sta07,BT14}. 

The understanding of the reason behind the very large thermal neutron capture cross sections with neutron poisons is not yet fully understood \cite{Cohen1971}, and most likely the phenomenon is a statistical one. A simple picture of a capture through an isolated compound nucleus resonance seems to require very stringent conditions on the resonance position, which are seldom met. Accordingly, with a wave of hands, nuclear physicists make the unavoidable assumption that these cross sections are large by fortunate or unfortunate chance! An attempt to get a more precise statement concerning this matter, was recently made in Ref. \cite{HCK2016}. In this reference it is suggested that the capture process proceeds through a simple 1p-2h doorway in the compound system of the type proposed in connection with intermediate structure \cite{BF1963, KRY1963, Feshbach1993, CH2013}. Though the large values of the cross section can be obtained with a judicial choice of the energy of the doorway resonance, the probability for this to actually happen was found to be \cite{HCK2016},
\begin{equation}
P(\eta_0) = \frac{1}{2\pi}\frac{1}{1 + \eta_{0}},
\end{equation}
where $\eta_{0} \equiv \Gamma_{D, n}/\Gamma_{q, n}$ is a measure of the doorway enhancement. $\Gamma_{D,n}$, and $\Gamma_{q,n}$ are the doorway and compound neutron widths, respectively. Generally, the former is in the keV region, while the latter in the eV region.
Therefore,  there is a very small probability for the occurrence of the doorway enhancement. Accordingly,
very large values of neutron capture cross sections are inhibited by statistics, and one is bound to remain in the random phenomenon domain! This conclusion would also hold for unstable nuclei. 

Recently, interest in extracting the neutron capture cross section by stable nuclei at higher energies through the (d,p) reaction has arisen in part for application to next generation reactors (fast breeder reactors fueled by $^{238}$U, and $^{232}$Th), and in part for the study of the reaction mechanism of weakly bound stable nuclei \cite{Moro1-2015}. There is also potential application to the production of elements in the r-process of nucleosynthesis. In a recent publication Ref. \cite{Potel2015}, tested the surrogate method \cite{Escher2012} in the case of (d,p) reaction on the actinide nuclei to be used in these projected reactors. For this purpose, they employed the theory of inclusive non-elastic breakup reactions, where the proton is treated as a spectator, merely scattering off the target, and the neutron is captured by the target, and, at higher energies, inelastically scattering from it. Other papers on the (d,p) reaction were also published recently dealing with the same issue \cite{Moro1-2015, Carlson2015}. Ref. \cite{Moro1-2015} also discussed the application of this  hybrid picture (direct breakup followed by compound nucleus formation of the subsystem)  to the reaction $^{6}$Li +$^{209}$Bi $\rightarrow \alpha + X$, at $E_{Lab.}$ = 24 MeV and 32 MeV. In \cite{Potel2016}, a comprehensive discussion of the (d, p) reactions along the above lines are given.

At much higher deuteron or other breaking projectiles energies, researchers relied on the very simple but physically transparent Serber model \cite{Serber1950}. The applications at low energies mentioned above rely on  theories of inclusive breakup reactions developed back in the late 70's and in the 80's \cite{Baur1978, Baur1980-1, Baur1980-2,UT1981, UT1984, IAV1985, HM1985, Austern1987}. So far, no attempt has been made to apply the hybrid theory to the extraction of the neutron capture cross section involving radioactive nuclei. Our purpose in this work is to supply this missing information, which can be of great importance to the nucleosynthesis of elements in the r-process where many radioactive isotopes are encountered along the way to the stable elements. 
\begin{table}
\begin{center}
\begin{tabular}
[c]{|l|l|l|l|}\hline
 Nucleus & Cross section (barn)  \\  \hline
$^9{\rm Be}$ &[8.77$ \pm 0.35] \times \text  10^{-3}$  \\
$^{\rm10}{\rm B}$ & 0.5$\pm$0.0.1 \\
$^{14}{\rm N}$ & [79.8$\pm1.4]\times \text 10^{-3}$ \\
$^{15}{\rm N}$ & [0.024$\pm0.008]\times \text10^{-3}$ \\
$^{16}{\rm O}$ & [0.19$\pm0.019 ]\times \text 10^{-3}$\\
$^{20}{\rm Ne}$&[37$\pm4]\times \text10^{-3}$\\
$^{21}{\rm Ne}$&0.666$\pm0.110$\\
$^{28}{\rm Si}$ & [177$\pm 5]\times \text10^{-3}$ \\
$^{40}{\rm Ar}$ & 0.660$\pm$0.01 \\
$^{40}{\rm Ca}$ & 0.41$\pm$ 0.02 \\
$^{56}{\rm Fe}$ & 2.59 $\pm$0.14 \\
$^{59}{\rm Co}$ & 37.18 $\pm$ 0.06 \\
$^{58}{\rm Ni}$ & 4.5$\pm$0.2 \\ 
$^{63}{\rm Cu}$ & 4.52$\pm$ 0.02 \\  \hline
\end{tabular}
\begin{tabular}
[c]{|l|l|l|l|}\hline
 Nucleus & Cross section (barn)  \\  \hline
$^{84}{\rm Kr}$ & 0.111$\pm$0.015 \\
$^{90}{\rm Zr}$ & 0.011$\pm$0.005 \\
$^{103}{\rm Rh}$ & 145$\pm$2 \\
$^{113}{\rm Cd}$ & [2.06$\pm0.04$]$\times \text 10^{4}$ \\
$^{114}{\rm Cd}$ & 0.34$\pm$0.02 \\
$^{135}${\rm Xe} & $2.65 \times 10^6$ \cite{Sta07}\\
$^{136}${\rm Xe} & $\sim 1 \times 10^{-3}$ \cite{BT14}\\
$^{149}{\rm Sm}$ & [4.014$\pm 0.06$]$\times \text 10^{4}$ \\
$^{157}{\rm Gd}$ & [2.54$\pm0.008$]$\times \text 10^{5}$ \\
$^{159}{\rm Tb}$ & 23.3$\pm$0.4 \\
$^{208}{\rm Pb}$ & [0.23$\pm0.03$]$\times 10^{-3}$ \\
$^{209}{\rm Bi}$ & 0.0338$\pm$0.0007 \\
$^{232}{\rm Th}$ & 7.35$\pm$0.03 \\
$^{238}{\rm U}$ & 2.68$\pm$ 0.019\\ \hline
\end{tabular}
\caption{Neutron capture cross section for several nuclei across the periodic table. The choice of the nuclei was dictated by the mass region and the disparity in the value of the thermal neutron capture cross section between adjacent nuclei or isotopes, when available. The full compilation can be found in \cite{Mughab2003}. Data for Xe isotopes were taken from Refs. \cite{Sta07,BT14}. }
\end{center}
\end{table}

\section{Surrogate reactions with neutron poisons}

Neutron poisons are nuclei having very large neutron reaction cross sections. Based on our discussion in the previous Sections, one also expects that the cross sections induced in (d,p) reactions would be also proportionally large. The study of neutron capture by radioactive nuclei is important both for the understanding the role of the excess neutrons on the capture process as well as for application to r-process nucleosynthesis, where the capture happens faster than the $\beta$-decay of the nuclei produced in core-collapse supernova.  Whereas the capture by stable nuclei can be studied in the laboratory using the usual arrangement of a fixed stable nuclear target, the case of radioactive nuclei, is more complicated. Depending on the lifetime of the nucleus, using it as a target can be possible, such as the case of $^{135}$Xe, but if the lifetime is in the millisecond or shorter, one has to use the capturing nucleus as a projectile and, the deuteron as the target. Because the reaction with a neutron poison is much larger than usual, we propose that it can be used to test the reliability of  surrogate method. Next we estimate of cross sections of (d,p) reactions on such nuclei that can be used a a guide for experimental studies.

\begin{figure}
\begin{center}
\includegraphics[scale=0.35]{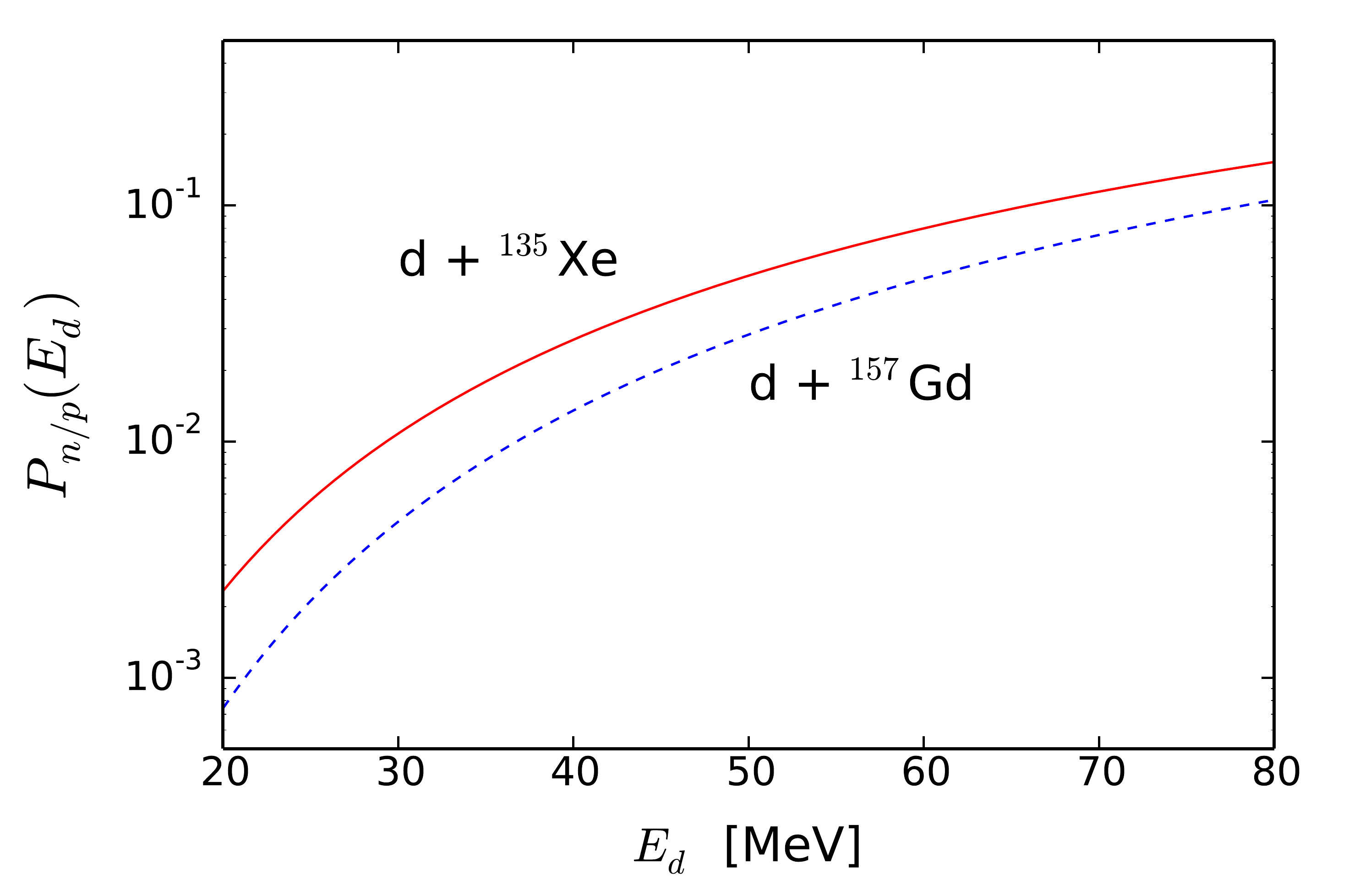}
\caption{Neutron surrogate factor for the reactions d + $^{135}$Xe and d + $^{157}$Gd as a function of the deuteron projectile energy $E_d$.} \label{f3}
\end{center}
\end{figure} 

In order to calculate the pertinent S-matrices needed for a numerical estimate, we define $S_i(b)=\exp{[i\chi_i(b)]}$ with $i=n,p$, in terms of eikonal phase $\chi_i(b)]$. In collisions at high energies, instead of Eq. \eqref{chibp} a standard expression for the eikonal phase, is given in terms of the nucleon-nucleon cross sections by
\begin{equation}
 \chi_i(b) = i\dfrac{\sigma_{NN}^{(i)}}{4\pi}\int_0^\infty dq\, q\,  {\rho_i}(q) {\rho_A}(q) J_0(qb) +\chi_C(b), \label{phasec}
\end{equation} 
where $\sigma_{NN}^{(i)}$ is the isospin averaged nucleon-nucleon cross section,  ${\rho_A}(q)$ is the Fourier transform of the (spherical) nuclear matter density, and ${\rho}_{d}(i)$ is that for the nucleon, and   $J_0$ is the cylindrical Bessel function. We assume a gaussian density for the matter distribution in the nucleon with a gaussian parameter $a=0.71$ fm. The isospin average cross sections are given by $\sigma_{NN}^{(n)}=(N\sigma_{pp}+Z\sigma_{np})/A$ and $\sigma_{NN}^{(p)}=(Z\sigma_{pp}+N\sigma_{np})/A$. In Eq. \eqref{phasec}, $\chi_C(b)$ is the Coulomb phase, calculated according to Eq. (20) of Ref. \cite{BG06}. To account for nuclear recoil due to  Coulomb repulsion, the eikonal phase  in Eq. \eqref{phasec} is calculated with the replacement $b \rightarrow b'=a+\sqrt{a^2 + b^2}$ where $a = Z_dZ_Te^2/\mu v^2$ is half the distance of closest approach in a head-on collision of the deuteron and the target.

\begin{figure}
\begin{center}
\includegraphics[scale=0.35]{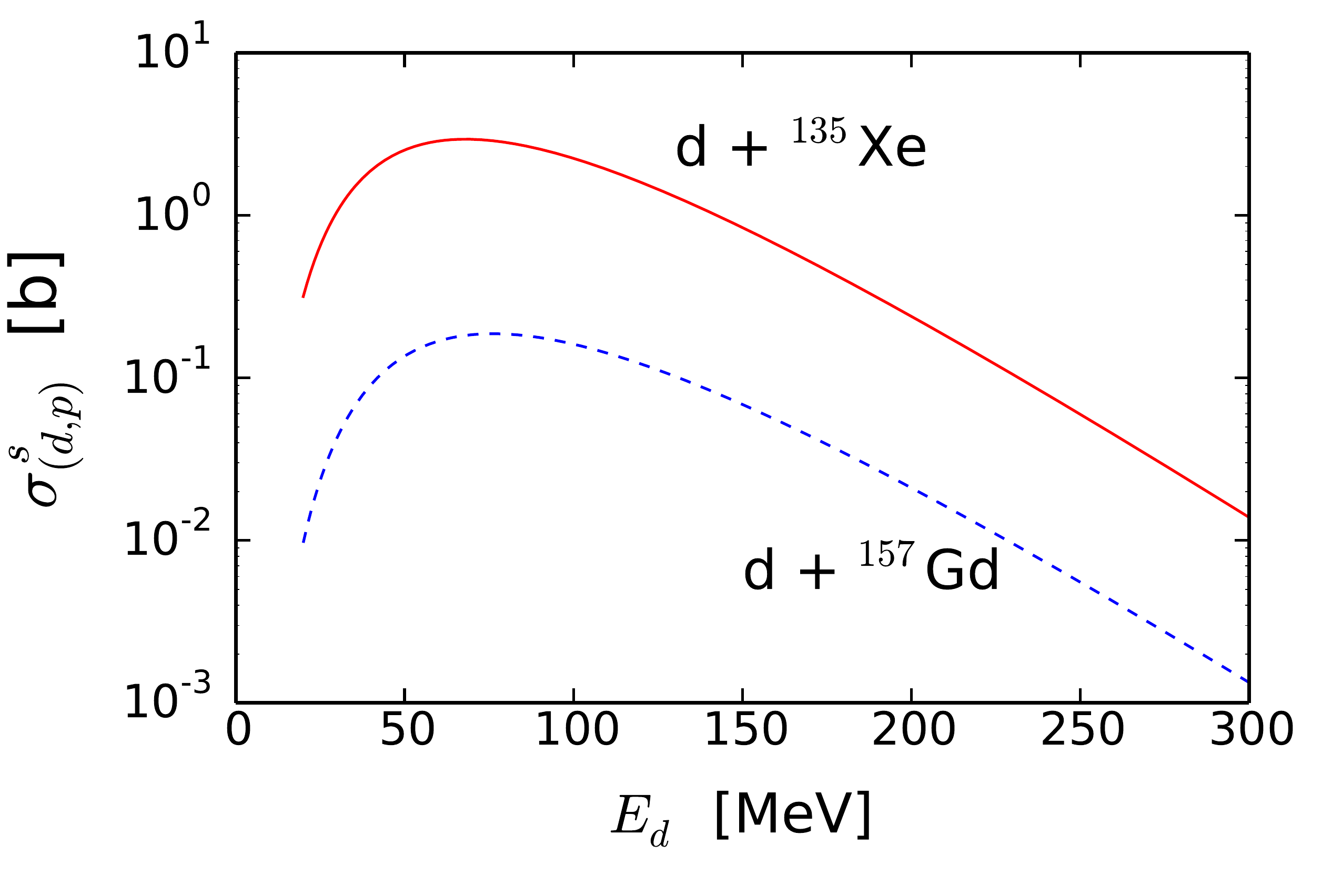}
\caption{Cross sections for thermal neutron capture in the  reactions d + $^{135}$Xe and d + $^{157}$Gd as a function of the deuteron energy $E_d$ in inverse kinematics.} \label{f4}
\end{center}
\end{figure}

In Figure \ref{f3} we plot the neutron surrogate factor for the reactions d + $^{135}$Xe and d + $^{157}$Gd as a function of the deuteron projectile energy $E_d$. We assume a s-wave deuteron wavefunction. One observes a steep increase of the surrogate factor which is mainly due to the Coulomb repulsion at small energies and a consequent reduced overlap between the deuteron wavefunction with the target in close collisions. Therefore, we see that the momentum matching function (see Figure \ref{f2}) and the neutron surrogate factors behave with opposite trends as a function of the projectile bombarding energy. As we show in Figure \ref{f4}, eventually, the momentum matching function wins over the surrogate factor energy dependence and the cross sections fall steeply with the projectile energy. In the Figure we plot the cross section in barns for the capture of thermal neutrons in (d,p) reactions as a function of the deuteron energy for a s-wave deuteron wave function. It is evident from the Figure that the best energy range for using (d,p) reactions as a tool for thermal neutron capture is in the region of $E_d =  50-150$ MeV. In this bombarding energy region the cross sections are very sizable. For d + $^{135}$Xe it reaches a few barns and for d + $^{157}$Gd it reaches a few hundreds of milibarns at their peak values.  These are large cross sections from the experimental point of view. In a usual radioactive beam facility such as in RIKEN/Japan, it would lead to a large number of events and very good statistics, allowing for a fine test of surrogate theories for nuclear reactions, such as in (d,p) reactions.

\begin{table}
\begin{center}
\begin{tabular}
[c]{|l|l|l|l|}\hline
 Nucleus & $\sigma^s$ (mb) & $\sigma^d$ (mb)  \\  \hline
$^{59}{\rm Co}$ & $3.62 \times 10^{-2}$ & $9.39 \times 10^{-3}$   \\
$^{58}{\rm Ni}$ & $4.11\times 10^{-3}$& $1.73 \times 10^{-3}$ \\ 
$^{63}{\rm Cu}$ & $4.24 \times 10^{-3}$ & $1.68 \times 10^{-3}$\\ 
$^{103}{\rm Rh}$ & $1.35\times 10^{-1}$ & $3.52 \times 10^{-2}$ \\
$^{135}${\rm Xe} & ${\bf 2.52 \times 10^3}$  & ${\bf 8.29 \times 10^2}$ \\ \hline 
\end{tabular}
\begin{tabular}
[c]{|l|l|l|l|}\hline
 Nucleus & $\sigma^s$ (mb)  & $\sigma^d$ (mb) \\  \hline
$^{149}{\rm Sm}$ & $35.5$ &9.74 \\
$^{157}{\rm Gd}$ & ${\bf 1.86 \times 10^2}$&${\bf 37.9}$ \\
$^{159}{\rm Tb}$ & $1.64\times 10^{-2}$&$3.04\times 10^{-3}$ \\
$^{232}{\rm Th}$ & $5.33\times 10^{-3}$&$1.36\times 10^{-3}$ \\
$^{238}{\rm U}$ & $1.65\times 10^{-3}$& $3.22 \times 10^{-4}$\\ \hline
\end{tabular}
\caption{Neutron induced thermal capture cross section in (d,p) reactions for several target nuclei and for 60 MeV deuteron projectiles. We highlight the very large cross sections for $^{135}$Xe and $^{157}$Gd. }
\label{tb2}
\end{center}
\end{table}

In Table \ref{tb2} we show our results for the cross sections for neutron induced thermal capture using (d,p) reactions for several nuclei and for 60 MeV deuterons in inverse kinematics. We highlight the very large cross sections for $^{135}$Xe and $^{157}$Gd. Such large cross sections make it amenable to experimental investigation using beams of  $^{135}$Xe and $^{157}$Gd incident on deuteron gas targets. On the other hand, our results also show that the zero point motion of the neutron inside the deuteron reduces the extracted capture cross section from its value for free neutrons by several orders of magnitudes.

\section{Discussion}

The calculations presented above refer to the proton yield in the (d,p) reaction corresponding to non-elastic breakup of the deuteron. The experimental yield will contain both the elastic breakup and the non-elastic breakup contributions. Accordingly, one is required to calculate the former and subtract it from the data in order to be able to compare to our estimates. The elastic breakup contribution is given by the following expression in the post-form representation,
\begin{equation}
\sigma_{EB} = \int d\Omega_{p}dE_{p}d\Omega_{n}dE_{n}\rho_{p}(E_{p})\rho_{n}(E_{n}) \left|\overline{\langle \chi_{p}^{(-)}\chi_{n}^{(-)}\left|V_{pn}\right|\chi^{(+)}_{d}\Phi_{d}\rangle}\right|^2,
\end{equation}
where $\overline{\langle \chi_{p}^{(-)}\chi_{n}^{(-)}\left|V_{pn}\right|\chi^{(+)}_{d}\Phi_{d}\rangle}$ is an average DWBA matrix element which represents the elastic breakup process,
$d + A \rightarrow p + n + A$. The density of continuum states of the proton and the neutron, are denoted by $\rho_{p}(E_p)$, and $\rho_{n}(E_n)$, respectively, (see paragraph below Eq. (\ref{IC})). The wave functions $\chi_{p}^{(-)}$, $\chi^{(-)}_{n}$ and $\chi_{d}^{(+)}$ are distorted waves of the proton, the neutron, and the incoming deuterons, respectively. These distorted waves are calculated with the corresponding optical potentials, $U_{p}$, $U_{n}$, and $U_{d}$. If one wishes to perform the calculation of $\sigma_{EB}$, using the prior-form representation, then it suffices to replace $V_{pn}$ by $U_{p} + U_{n} - U_{d}$ in the matrix element. Once $\sigma_{EB}$ is calculated, then the reduced data, 
$\sigma_{Exp} - \sigma_{EB}$ is constructed and can be compared to our results above.

\section{Conclusions}

In this paper we have supplied the theoretical framework needed to make credible estimates of the neutron capture cross section involving both stable and radioactive nuclei in inverse kinematics. The theoretical basis of our method is the theory of inclusive breakup reactions which has been developed in the 80's and used for the (d,p) reactions in 2015. 

The Serber-like model developed here was used to obtain the deuteron non-elastic breakup cross section for the reactions $^{135}Xe$(d, p)$^{136}$Xe and $^{157}$Gd(d,p)$^{158}$Gd. The calculation is performed at deuteron energies and appropriate cuts in the cross section formula which guarantees that the neutron capture cross section embedded in our formulae, corresponds to thermal neutron energies. The obtained deuteron cross sections for the above systems was found to be a few barns, making its measurement feasible. 

In free space the thermal neutron capture by $^{135}$Xe is measured by a cross section of the order of 2.5 x $10^{6}$ barns while that of $^{157}$Gd is 2.4 x $10^{5}$ barns. In the (d,p) reactions the extracted capture cross should in principle give similar results. This would require paying special attention to the effects of
zero point motion of the neutron inside the deuteron  as one performs an unfolding of our formulae. This conveys the message that the extracted capture cross section must be judicially corrected in order to obtain the desired "free" neutron capture one. The Surrogate method, being based on the same theory of inclusive breakup, should be similarly assessed in light of what we have obtained here \cite{Potel2015}.
\appendix

\section{Cross section dependence on the neutron momentum distribution}
\subsection{Theory for A(\textsc{d},p)B reactions}

In this Appendix we demonstrate a  relation between the (d,p) cross section to the neutron induced cross section emphasizing the dependence on the momentum distribution of nucleons in the nuclei. The c.m. recoil momentum of the spectator $p$ is denoted by ${\bf P}_f$ which is the negative of the recoil momentum of the $A+n=B$ system. If we denote the momentum transfer, which is the difference between the final and initial momenta, by ${\bf q}_{k}$, for the k-particle where $k=p,~n,~A$, we have
\begin{align}
{\bf q}_p  & ={\bf P}_f-\frac{m_p}{m_p+m_n}{\bf P}_i \simeq {\bf P}_f-{{\bf P}_i \over 2} \label{2.a}\\
{\bf q}_n  & =-\frac{m_n}{m_A+m_n}{\bf P}_f-\frac{m_n}
{m_p+m_n}{\bf P}_i\simeq -\frac{m_n}{m_B}{\bf P}_f-{{\bf P}_i\over 2}\label{2.b}\\
{\bf q}_A  & =-\frac{m_A}{m_B}{\bf P}_f+{\bf P}
_{i}\ .\label{2.c}
\end{align}
Because the mass arrangements are not the same in the initial and the final states, the reduced masses $\mu_{i}$ and $\mu_{f}$ associated with the initial and the final momenta are different. $\mu_{i}$ and $\mu_{f}$ are given by
\begin{equation}
\frac{1}{\mu_i}=\frac{1}{m_p+m_n}+\frac{1}{m_A}\simeq {1\over 2m_p}+{1\over m_A}
,\ \ \ \ \ \ \ \ \ \ \ \ \ \ \ \ \ \ \frac{1}{\mu_f}=\frac{1}{m_p}
+\frac{1}{m_B}\ .\label{3.3}
\end{equation}

The total energy $E$ is given by (we use $c=1$)
\begin{equation}
E=m_A+m_d+\frac{{\bf P}_{i}^{2}}{2\mu_{i}}-\epsilon_{np}=m_p+m_B+\frac{{\bf P}_{f}^{2}}{2\mu_{f}}+\epsilon_{nA}\label{6.4a}
\end{equation}
where $P_{i}^{2}/2\mu_{i}$ and $P_{f}^{2}/2\mu_{f}$ are the initial and final kinetic energies, and both $\epsilon_{np}$ and $\epsilon_{nA}$ are positive quantities denoting the binding energies of $n+p=d$ and the continuum energy of the $n+A=B$ system. We assume that $n+A$ is unbound.

The $Q$-value for the reaction is given by $ Q={{\bf P}_f^{2}}/{2\mu_f}+\epsilon_{nA}-{{\bf P}_i^2}/{2\mu_i}+\epsilon_{np}=\Delta m$, where $\Delta m= m_A+m_d-m_B-m_p$. The center of mass of particles ($n+p$) and ($n+A$) are given by 
${\bf R}_{np}={{m_p}{\bf r}_p+m_n{\bf r}_n}/{m_d}$, and  ${\bf R}_{nA}={m_n{\bf r}_{n}+m_A{\bf r}_A}/{m_B}$,
respectively. Since the center of mass remains fixed, the independent sets of coordinates are both the relative coordinates of the particles in the bound systems and the differences between the coordinates of the free particle and the center of mass of these particles. Thus,
${\bf r}_{np}   ={\bf r}_n-{\bf r}_p$, with ${\bf r}_{i}={\bf r}_A-{\bf R}_{np}$, and
${\bf r}_{nA}   ={\bf r}_n-{\bf r}_A$, with ${\bf r}_{f}={\bf r}_p-{\bf R}_{nA}$,
are the set of independent coordinates initially and finally, with the condition
$
m_p{\bf r}_p+m_n{\bf r}_n+m_A{\bf r}_A=0.
$
That is, the center of mass of the whole system is at rest.

The momenta associated with these coordinates can be calculated by using the definition of canonical momentum, i.e., ${\bf P}_{k}=\left(  1/i\right)\partial/\partial{\bf r}_{k}$, with $k=i$, $f$, $np$, $nA$ (for now, to simplify notation, we
use $\hbar=1$). We obtain
\begin{align}
{\bf p}_{np}  & =\mu_{np}\left(  \frac{{\bf p}_p}{m_p} -\frac{{\bf p}_n}{m_n}\right)  ,\ \ \ \ \ \ \ \ \ \ \ \ {\bf P}_{i}=\mu_{i}\left(  \frac{{\bf p}_p+{\bf p}_n}{m_d} -\frac{{\bf p}_A}{m_A}\right)  \ ,\nonumber\\
{\bf p}_{nA}  & =\mu_{nA}\left(  \frac{{\bf p}_n}{m_n}-\frac{{\bf p}_A}{m_A}\right)  ,\ \ \ \ \ \ \ \ \ \ \ \ {\bf P}_{f}=\mu_{f}\left(  \frac{{\bf p}_p}{m_p}-\frac{{\bf p}_n+{\bf p}_A}{m_B}\right)  \ ,
\end{align}
where $\mu_{np}$ and $\mu_{nA}$ are the reduced masses of ($n+p$) and ($n+A$), respectively. The kinetic energy operator, $T_{\mathrm{cm}}$, in the center of mass system is therefore
\begin{equation}
T_{\mathrm{cm}}=\frac{{\bf p}_{np}^{2}}{2\mu_{np}}+\frac{{\bf P}_{i}
^{2}}{2\mu_{i}}=\frac{{\bf p}_{nA}^{2}}{2\mu_{nA}}+\frac{{\bf P}_{f}^{2}}{2\mu_{f}}\ .\label{6.9}
\end{equation}

The total Hamiltonian for the system is
\begin{equation}
H=T_{\mathrm{cm}}+V_{np}+V_{pA}+V_{nA}\ ,\label{6.10}
\end{equation}
and the wave functions for the states of the initial ($n+p$)-bound and final ($n+A$)-unbound systems are
\begin{align}
\left(  \frac{{\bf p}_{np}^{2}}{2\mu_{np}}+V_{np}\right)  \phi_{np}\left({\bf r}_{np}\right)    =-\epsilon_{np}\phi_{np}\left(  {\bf r}_{np}\right), \ \ \ \ \ \ {\rm and }  \ \ \ \ \ 
\left(  \frac{{\bf p}_{nA}^{2}}{2\mu_{nA}}+V_{nA}\right)  \phi_{nA}\left({\bf r}_{nA}\right)    =\epsilon_{nA}\phi_{nA}\left(  {\bf r}_{nA}\right)  \ ,\label{6.12}
\end{align}
where both $\epsilon_{np}$ and $\epsilon_{nA}$ are positive quantities. We denote by $\chi_{i}\left(  {\bf r}_{i}\right)  $ and $\chi_{f}\left({\bf r}_{f}\right)  $\ the wavefunctions of the free particle $A$ with respect to the center of mass of ($n+p$) in the initial state, and of $p$ with respect to the center of mass of $(n+A)$ in the final state, respectively. The total wavefunctions $\psi\left(  {\bf r}_{np},{\bf r}_{i}\right)  $ and $\psi\left(  {\bf r}_{nA},{\bf r}_{f}\right)  $ can be expanded as follows
\begin{equation}
\psi\left(  {\bf r}_{np},{\bf r}_{i}\right)  =\sum_{\alpha}\phi_{\alpha}\left(  {\bf r}_{np}\right)  \chi_{\alpha}\left(  {\bf r}_{i}\right)
,\ \ \ \ \ \ \ \ \ \ \ \ \ \ \psi\left(  {\bf r}_{nA},{\bf r}_{f}\right)  =\sum_{\alpha}\phi_{\alpha}\left(  {\bf r}_{nA}\right)
\chi_{\alpha}\left(  {\bf r}_{f}\right)  \ ,\label{6.13}
\end{equation}
where $\alpha$ stands for a complete set of states of the bound system and the free particle. From Eqs. (\ref{6.13}) we get
\begin{eqnarray}
\chi_{i}\left(  {\bf r}_{i}\right)
=\int\phi_{np}^{\ast}\left( {\bf r}_{np}\right)  \psi\left(
{\bf r}_{np},{\bf r}_{i}\right) d^{3}r_{np}, \ \ \ \ \ \ \ \ {\rm and} \ \ \ \  \ \ 
\chi_{f}\left( {\bf r}_{f}\right) = \int\phi_{nA}^{\ast}\left(  {\bf r}_{nA}\right)  \psi\left(  {\bf r}_{nA},{\bf r}_{f}\right)  d^{3}r_{nA}\ .\label{6.14}
\end{eqnarray}

To obtain the equation satisfied by $\chi_{i}\left(  {\bf r}_{i}\right),$\ we note that $P_{f}^{2}=-\nabla_{f}^{2}$. From (\ref{6.4a}), (\ref{6.9}) and (\ref{6.10}) we have
\begin{align}
H\psi\left(  {\bf r}_{nA},{\bf r}_{f}\right)    =\left(  \frac{{\bf p}_{nA}^{2}}{2\mu_{nA}}+\frac{{\bf P}_{f}^{2}}{2\mu_{f}}+V_{np}+V_{pA}+V_{nA}\right)  \psi\left(  {\bf r}_{nA},{\bf r}_{f}\right) 
 =\left(  \frac{{\bf P}_{f}^{2}}{2\mu_{f}}+\epsilon_{nA}\right)\psi\left(  {\bf r}_{nA},{\bf r}_{f}\right)  \ ,
\end{align}
or
\begin{equation}
\frac{1}{2\mu_{f}}\left(  \nabla_{f}^{2}+P_{f}^{2}\right)  \psi\left({\bf r}_{nA},{\bf r}_{f}\right)  =\left(  \frac{{\bf p}_{nA}^{2}}{2\mu_{nA}}+V_{np}+V_{pA}+V_{nA}-\epsilon_{nA}\right)  \psi\left(
{\bf r}_{nA},{\bf r}_{f}\right)  \ .\label{6.15a}
\end{equation}
Applying the operator $\left(  \nabla_{f}^{2}+P_{f}^{2}\right)  $ to $\chi _{f}\left(  {\bf r}_{f}\right)  $ in Eq. (\ref{6.14}) and using Eqs. \eqref{6.12} and (\ref{6.15a}), we obtain
\begin{equation}
\left(  \nabla_{f}^{2}+P_{f}^{2}\right)  \chi_{f}\left(  {\bf r}_{f}\right)  =2\mu_{f}\int\phi_{nA}^{\ast}\left(  {\bf r}_{nA}\right)\left[  V_{np}+V_{pA}\right]  \psi\left(  {\bf r}_{nA},{\bf r}_{f}\right)  d^{3}r_{nA}\ .\label{6.15b}
\end{equation}

The solution for $\chi_{f}\left(  {\bf r}_{f}\right)  $\ can be obtained by using the outgoing Green's function $
-{\exp\left[  iP_{f}\left\vert {\bf r}_{f}-{\bf r}\right\vert\right]  }(4\pi\left\vert {\bf r}_{f}-{\bf r}\right\vert )^{-1}$.
Thus,
\begin{equation}
\chi_{f}\left(  {\bf r}_{f}\right)  =-\frac{\mu_{f}}{2\pi}\int\frac{\exp\left[  iP_{f}\left\vert {\bf r}_{f}-{\bf r}\right\vert \right]}{4\pi\left\vert {\bf r}_{f}-{\bf r}\right\vert }\phi_{nA}^{\ast}\left({\bf r}_{nA}\right)  \left[ V_{np}+V_{pA}\right]  \psi\left({\bf r}_{nA},{\bf r}\right)  \ d^{3}r_{nA}\ d^{3}r_f\ ,\label{6.16}
\end{equation}
and the reaction scattering amplitude is
\begin{equation}
f\left(  \theta,\phi\right)  =-\frac{\mu_{f}}{2\pi}\int\phi_{nA}^{\ast}\left({\bf r}_{nA}\right)  \ \exp\left(  -i{\bf P}_{f}\cdot{\bf r}_{f}\right)  \ \left[  V_{np}+V_{pA}\right]  \ \psi\left(  {\bf r}_{nA},{\bf r}_{f}\right)  \ d^{3}r_{nA}\ d^{3}r_{f}\ .\label{fidr0}
\end{equation}

The above result is exact. No approximations have been made so far.

\subsection{Serber \& Butler model}

As it stands, Eq. \eqref{fidr0} is not very useful. Here we will use the Born approximation (also called Serber or Butler theory \cite{Serber1950,But50} for direct reactions) which has been very successful to describe direct reactions. It is usually applicable to the situation in which $n+A$ is a bound system. In the Butler model one assumes that due to the short range of the $V_{np}$ and $V_{pA}$ interactions (neglecting Coulomb), the integrals in Eq.  \eqref{fidr0} selects nearly equal coordinates of the wavefunctions involved. Therefore, one may replace $\psi\left(  {\bf r}_{nA},{\bf r}_{f}\right) \simeq \psi\left(  {\bf r}_{np},{\bf r}_{i}\right)$ and assume a plane wave for the relative motion to leading order. That is,
\begin{equation}
\psi\left(  {\bf r}_{nA},{\bf r}_{f}\right) \simeq \phi_{np}\left(  {\bf r}_{np}\right)  \exp\left[  i{\bf P}_{i}\cdot{\bf r}_{i}\right]  = \phi_{np}\left(  {\bf r}_{np}\right)  \exp\left[  -i{\bf P}_{i}\cdot{\bf R}_{np}+i{\bf P}_{i}\cdot{\bf r}_A\right]   \ ,
\end{equation}
and the reaction scattering amplitude becomes
\begin{align}
f\left(  \theta,\phi\right)   & \simeq-\frac{\mu_{f}}{2\pi}\int\phi_{nA}^{\ast}\left(  {\bf r}_{nA}\right)  \ \exp\left(  -i{\bf P}_{f}\cdot{\bf r}_{f}\right)  \left[  V_{np}+V_{pA}\right]  \exp\left[  i{\bf P}_{i}\cdot{\bf r}_{i}\right]  \ \phi_{np}\left(  {\bf r}_{np}\right)  \ d^{3}r_{nA}\ d^{3}r_{f}\ \nonumber\\
& =-\frac{\mu_{f}}{2\pi}\int\phi_{nA}^{\ast}\left(  {\bf r}_{nA}\right)\exp\left(  i{\bf q}_n\cdot{\bf r}_{nA}-i{\bf q}_p\cdot{\bf r}_{np}\right)  \left[  V_{np}+V_{pA}\right] 
\phi_{np}\left(  {\bf r}_{np}\right)  \ d^{3}r_{np}\ d^{3}r_{nA}\ ,\label{fidr1}
\end{align}
where we made use of the equalities
\[
-{\bf P}_{f}\cdot{\bf r}_{f}+{\bf P}_{i}\cdot{\bf r}_{i}={\bf q}_n\cdot{\bf r}_{nA}-{\bf q}_p\cdot{\bf r}_{np},\ \ \ \ \ \ \ \ \mathrm{and}\ \ \ \ d^{3}r_{nA}\ d^{3}r_{f}=d^{3}
r_{np}\ d^{3}r_{nA}\ .
\]

The transfer reaction cross section, $\sigma\left(  \theta,\phi\right) = (v_f/v_i)|f\left(  \theta,\phi\right)|^2$,  is thus given by
\begin{align}
\sigma\left(  \theta,\phi\right)   & =\frac{\mu_{i}\mu_{f}}{(2\pi)^{2}}\frac{P_{f}}{P_{i}}\left\vert \int\phi_{nA}^{\ast}\left(  {\bf r}_{nA}\right)  \exp\left(  i{\bf q}_n\cdot{\bf r}_{nA}-i{\bf q}_p\cdot{\bf r}_{np}\right)  \left[  V_{np}+V_{pA}\right] \phi_{np}\left(  {\bf r}_{np}\right)  \ d^{3}r_{np}
\ d^{3}r_{nA}\right\vert ^{2}.\label{sigdir1}
\end{align}

Eqs. (\ref{fidr1}) and (\ref{sigdir1}) can be further simplified by neglecting the interaction $V_{pA}$\ between particles $p$ and $A$. This is done because in the reaction $(n+p)+A\rightarrow p+(n+A)$ particles $A$ and $p$ never appear in a bound or resonant state. Hence, the integrals factorize, and we get
\begin{align}
\sigma\left(  \theta,\phi\right)   & \simeq\frac{\mu_{i}\mu_{f}}{(2\pi)^{2}}\frac{P_{f}}{P_{i}}\left\vert \int\exp\left(  i{\bf q}_n\cdot
{\bf r}_{nA}\right)  \ \phi_{nA}^{\ast}\left(  {\bf r}_{nA}\right)\ d^{3}r_{nA}\int\exp\left(  -i{\bf q}_p\cdot{\bf r}_{np}\right)V_{np}\left(  {\bf r}_{np}\right)  
\ \phi_{np}\left(  {\bf r}_{np}\right)  \ d^{3}r_{np}\right\vert ^{2}\label{sigdir2}\ .
\end{align}

Now, using the identity $ \left[  -{\nabla_{np}^{2}}/{(2\mu_{np})}-{q_p^{2}}/{(2\mu_{np})}\right]\exp\left(  -i{\bf q}_p\cdot{\bf r}_{np}\right)  =0$,
and the first of Eq. (\ref{6.12}) it is straightforward to show that
\begin{align}
 \int\exp\left(  -i{\bf q}_p\cdot{\bf r}_{np}\right)  V_{np}\left({\bf r}_{np}\right)  \ \phi_{np}\left(  {\bf r}_{np}\right)
\ d^{3}r_{np}
 =-\left(  \epsilon_{np}+\frac{q_p^{2}}{2\mu_{np}}\right)  \int\exp\left(-i{\bf q}_p\cdot{\bf r}_{np}\right)  \ \phi_{np}\left(  {\bf r}
_{np}\right)  \ d^{3}r_{np}\ .\label{intex}
\end{align}

Thus,
\beq
f\left(  \theta,\phi\right)    \simeq-\frac{\mu_{f}}{2\pi}\left(\epsilon_{np}+\frac{q_p^{2}}{2\mu_{np}}\right)  \mathcal{G}_{np}\left({\bf q}_p\right)  \mathcal{G}_{nA}\left(  {\bf q}_n\right),
\eeq
and
\beq
\sigma\left(  \theta,\phi\right)    \simeq\frac{\mu_{i}\mu_{f}}{(2\pi)^{2}
}\frac{P_{f}}{P_{i}}\left(  \epsilon_{np}+\frac{q_p^{2}}{2\mu_{np}}\right)^{2}\left\vert \mathcal{G}_{np}\left(  {\bf q}_p\right)  \right\vert^2 \left\vert\mathcal{G}_{nA}\left(  {\bf q}_n\right)  \right\vert ^{2}\ ,
\eeq
where
\begin{align}
\mathcal{G}_{np}\left(  {\bf q}_p\right)    =\int\exp\left(-i{\bf q}_p\cdot{\bf r}_{np}\right)  \ \phi_{np}\left(  {\bf r}_{np}\right)  \ d^{3}r_{np} \ \ \ \ \ \ \ {\rm and} \ \ \ \ \ \mathcal{G}_{nA}\left(  {\bf q}_n\right)    =\int\exp\left(i{\bf q}_n\cdot{\bf r}_{nA}\right)  \ \phi_{nA}^{\ast}\left({\bf r}_{nA}\right)  \ d^{3}r_{nA}\ .\label{g1g2}
\end{align}

We thus have shown that the transfer cross section samples momentum distributions of the spectator (or the participant) within the $d=n+p$ nucleus. It also shows that the same information is extracted for the target. This is manifested in the factorization of the form factors in two separated parts, one involving only $n$ and $p$, and the other involving $n$ and $A$.  The two integrals in Eq. \eqref{g1g2} are however constrained through the energy-momentum conservation relations Eq. (\ref{2.a}-\ref{2.c},\ref{6.4a}) which need to be enforced.

\subsection{Relation to the neutron-nucleus cross section}

Following the discussion above, we will show below that the A(d,p) reaction can be related to the free n-A cross section by means of 
\begin{equation}
{d\sigma_{dp} \over dE_p d\Omega_p}   = \frac{\mu_{i}\mu_{f}}{16\mu_{nA}^2}\frac{P_{f}}{P_{i}}\left(  \epsilon_{np}+\frac{q_p^{2}}{2\mu_{np}}\right)^{2}\left\vert \mathcal{G}_{np}\left(  {\bf q}_p\right)  \right\vert^2 {d\sigma_{nA} \over dE_{nA} d\Omega_{nA}}\ , \label{sigen}
\end{equation}
But notice that we have introduced the notation $E_{nA}$ and $\Omega_{nA}$ so that it becomes clear that the energies and scattering angles for $nA$ scattering and that of the emitted proton in the (d,p) reaction, e.g., $E_p$ and $\Omega_p$ are related by the energy-momentum conservation relation Eq. (\ref{2.a}-\ref{2.c},\ref{6.4a}). It also means that the reaction will sample only those values of ${\bf q}_p$ in accordance with the energy-momentum conservation relations Eq. (\ref{2.a}-\ref{2.c},\ref{6.4a}). 

For large projectile energies the energy-momentum conservation {\it will sample very large values of ${\bf q}_p$ in the deuteron form factor} $\mathcal{G}_{np}\left(  {\bf q}_p\right)$, largely reducing  $\left\vert \mathcal{G}_{np}\left(  {\bf q}_p\right)  \right\vert^2$ and the (d,p) cross section. This is because the proton almost at rest in the deuteron will need a large momentum transfer to become a free energetic proton in the final channel. In the other end of the spectrum, if the final proton has a small energy, the momentum transfer ${\bf q}_p$ will be small and $\left\vert \mathcal{G}_{np}\left(  {\bf q}_p\right)  \right\vert^2$ might be large. But then the neutron energy has to be accordingly large and the cross section  $d\sigma_{nA}/dE_{nA} d\Omega_{nA}$ will become very tiny. Therefore, we will expect that the (d,p) cross section in Eq. \eqref{sigen} will, for any value of ${\bf q}_p$, be much smaller than that for n-A. Since the later is extremely large for neutron poisons such as $^{135}$Xe, we expect that  the $^{135}$Xe(d,p) cross section might still be very large for some range of incident energies and phase space combinations.

\subsection{Regularization of neutron transfer to the continuum}

While $\mathcal{G}_{np}$ converges because $\phi_{np}\left(  {\bf r}_{np}\right)$ is a bound state, $\mathcal{G}_{nA}$ diverges due to $\phi_{nA}\left(  {\bf r}_{nA}\right)$ being a continuum wavefunction. An evaluation of the second  integral  in Eq. \eqref{g1g2} needs a regularization procedure, which can be achieved by working with the asymptotic behavior of the scattering wavefunction when the energy is close to a resonance.  We will assume that $\phi_{nA}\left(  {\bf r}_{nA}\right)$ represents a resonant state and that the resonance energy is given by $E_0 + i\Gamma/2$, so that $k^2/2\mu_{nA}=E_0 + i\Gamma/2$  and that asymptotically the system has obeys the outgoing boundary condition  $\phi_{nA} \rightarrow f_{nA}(\theta_{n})\exp(iq_{n} r)/r$. 

We assume that close to the resonant momentum $k$ the form factor behaves as
\begin{equation}
 \mathcal{G}_{nA}\left(  {\bf q}_n\right) = {\sqrt{C}g_{n}({\bf k}) \over q_n^2 - k^2  - i\epsilon_{0^+}} \ ,
\label{modis1}
\end{equation}
where $C$ is a constant to enforce correct dimensionality, $\epsilon_{0^+}$ is a small positive number to enforce outgoing spherical conditions, and we look for a function $g_{n}({\bf k})$ which satisfies it.  The inverse Fourier transform using the asymptotic limit of $ \phi_{nA}$ yields
\begin{eqnarray}
 f_n(\theta_n){\exp(iq_n r)\over r} &=& {1\over (2\pi)^3} \int d^3 k \exp\left(
i{\bf k}\cdot{\bf r}_{nA}\right) \mathcal{G}_{nA}\left(  {\bf q}_n\right) =  {1\over (2\pi)^3}\int d^3k {\sqrt{C}g_{n}({\bf q}_n) \over q_n^2 - k^2 -i\epsilon_{0^+}} \exp\left(i{\bf k}\cdot{\bf r}_{nA}\right) \nonumber \\
&=& {\sqrt{C}g_{n}({\bf q}_n)\over 2\pi^2 }\int dk {kj_0(kr) \over q_n^2 - k^2 -i\epsilon_{0^+}} \simeq {i\over \pi} \sqrt{C}g_{n}({\bf q}_n){\exp(iq_nr)\over r}\ .
\label{modis}
\end{eqnarray}
In the last step we have made use that $r$ is very large, followed by the theorem of residues to calculate the remaining integral.

The above result shows that, close to a resonance, $g_{n}({\bf q}_n)=-i\pi f(\theta_n)/\sqrt{C}$. Therefore, the cross section in Eq.  factorizes as
\begin{equation}
\sigma\left(  \theta,\phi\right)    \simeq\frac{\mu_{i}\mu_{f}}{16\mu_{nA}^2}\frac{P_{f}}{P_{i}}\left(  \epsilon_{np}+\frac{q_p^{2}}{2\mu_{np}}\right)^{2}\left\vert \mathcal{G}_{np}\left(  {\bf q}_p\right)  \right\vert^2 {d\sigma_{nA} \over dE_n d\Omega_n}\ , \label{sigen}
\end{equation}
where
\begin{equation}
 {d\sigma_{nA} \over dE_n d\Omega_n}={\left| f_{n}(\theta_n) \right|^2/C\over (E-E_0)^2+\Gamma^2/4} \ ,
\end{equation}
where for a given energy,  $d\sigma_{nA} / d\Omega_n =\left| f_{n}(\theta_n) \right|^2$. Although this result has been derived by assuming a resonance behavior of the neutron-nucleus cross section, one can use it for the very large cross sections for neutron scattering off nuclei at very low energies.

\section*{Acknowledgment}
CAB acknowledges support from the U.S. NSF Grant number 1415656 and the U.S. DOE Grant number DE-FG02-08ER41533, and by the Fulbright U.S. Scholar Program.


\begin{thebibliography}{99}
\bibitem{BBF57} E.M. Burbidge, G.R. Burbidge, W.A. Fowler, and F. Hoyle, Rev. Mod. Phys. 29, 547 (1957).
\bibitem{BG10} C.A. Bertulani and A. Gade, Phys. Rep. 485, 195 (2010). 
\bibitem{BK16} C.A. Bertulani and T. Kajino, Prog. Part. Nucl. Phys. 89, 56 (2016). 
\bibitem{CB70} J.D. Cramer and H.C. Britt, Phys. Rev. C 2, 2350 (1970); Nucl. Sci. Eng. 41, 177 (1970).
\bibitem{ED06} J. Escher and F.S. Dietrich, Phys. Rev. C 74 054601 (2006).
\bibitem{Fors07} C. Forss\'en, F. S. Dietrich, J. Escher, R. D. Hoffman, and K. Kelley, Phys. Rev. C 75, 055807 (2007).
\bibitem{DJ15} Q. Ducasse , B. Jurado, et al., arXiv:1512.06334  (2015).
\bibitem{BJ11} G. Boutoux, B. Jurado, et al., EPJ Web of Conferences 17, 06005 (2011).
\bibitem{Burk06} J.T. Burke et al., Phys. Rev. C 73, 054604 (2006).
\bibitem{TH07} F. Tovensson and T. S. Hill, Phys. Rev. C 75, 034610 (2007).
\bibitem{Bas09} M S Basunia et al, Nucl. Instrum. Methods B 267, 1899 (2009).
\bibitem{Alm09} J. M. Allmond et al., Phys. Rev. C 79, 054610 (2009).
\bibitem{Nay14} B.K. Nayak, Pramana: J.  Phys. 83, 799 (2014). 
\bibitem{GrassSand67} P. Grassberger and W. Sandhas Nucl. Phys. {B2}, 181 (1967).
\bibitem{AGS67} E.O. Alt, P. Grassberger, and W. Sandhas, Nucl. Phys. B2, 167 (1967).
\bibitem{MES12} A. M. Mukhamedzhanov, V. Eremenko and A. I. Sattarov, Phys. Rev. C 86, 034001 (2012).
\bibitem{UT1981}T. Udagawa and T. Tamura, Phys. Rev. C 24, 1348 (1981).
\bibitem{UT1984}T. Udagawa, X. H. Li, and T. Tamura, Phys. Lett. B  135 , 333(1984).
\bibitem{IAV1985}M. Ichimura, N. Austern, and C. M. Vincent, Phys. Rev. C  32,431 (1985).
\bibitem{HM1985}M. Hussein and K. McVoy, Nucl. Phys. A  445, 124 (1985).
\bibitem{Austern1987}N. Austern, Y. Iseri, M. Kamimura, M. Kawai, G. Rawitscher,and M. Yahiro, Phys. Rep. 154, 125 (1987).
 \bibitem{HFM1990}M. Hussein, T. Frederico, and R. Mastroleo, Nucl. Phys. A 511, 269 (1990).
\bibitem{Ichimura1990}M. Ichimura, Phys. Rev. C 41, 834 (1990).
\bibitem{Moro1-2015}J. Lei and A. M. Moro, Phys. Rev. C 92, 044616 (2015).
\bibitem{Moro2015}J. Lei and A. M. Moro, C 92, 061602(R) (2015).
\bibitem{HM1989} M. S. Hussein and R. C. Mastroleo, Nucl. Phys. A 491, 468 (1989)
\bibitem{BM1992} C. A. Bertulani and K. W. McVoy Phys. Rev. C 46, 2638 (1992). 
\bibitem {Canto1998} L. F. Canto, R. Donangelo, L. Mattos, M. S. Hussein, and P. Lotti, Phys. Rev. C 58, 1107 (1998). 
\bibitem{ABH92} A.N.F. Aleixo, C.A. Bertulani and M.S. Hussein, Phys. Rev. C 45, 2403 (1992).
\bibitem{HM85} M. S. Hussein and K. W. McVoy, Nucl. Phys. A 445, 124 (1985). 
\bibitem{Hul57} L. Hulth\'en and M. Sugawara, in Handbuch der Physik, edited by S. Flugge (Springer-Verlag, Berlin, 1957), Vol. 39.
\bibitem{MHE87} R. Machleidt, K. Holinde, and Ch. Elster, Phys. Rep. 149, 1 (1987).
\bibitem{Lac81} M.Lacombe, et al, Phys. Lett. B 101,139 (1981).
\bibitem{Eli12} E. Piasetzky, J. Phys.: Conf. Series 381, 012005 (2012).
\bibitem{Hen14} O. Hen, B.A. Li, W.J. Guo, L.B. Weinstein, and E. Piasetzky, Phys. Rev. C 91 (2), 025803 (2014).
\bibitem{YF13} Yoshiaki Yura and  Yusei Fujita, Oral Science International 10, 9 (2013).
 \bibitem{Nov13} N. Dewi et al., Biomedicine \& Pharmacotherapy, 67, 451 (2013).
 \bibitem{Mughab2003} S. F. Mughabghab, ``Thermal Neutron Capture Cross Sections Resonance Integrals and G-Factors", Int. Atomic. Energy Agency, INDC(NDS)-440 (2003).
\bibitem{Sta07} W. M. Stacey, ``Nuclear Reactor Physics", 2nd Ed. (Wiley-VCH, 2007), p. 213.
\bibitem{BT14} Megha Bhike and W. Tornow, Phys. Rev. C 89, 031602(R) (2014).
\bibitem{Cohen1971} B. L. Cohen, ``Concepts of Nuclear Physics", McGraw-Hill, Inc. (1971).
\bibitem{HCK2016} M. S. Hussein, B. V. Carlson, and A. K. Kerman, Acta. Physica Polonica B (2016);  arXiv:1506.03246
\bibitem{BF1963} B. Bloch and H. Feshbach, Ann. Phys. (NY), 23, 47 (1963).
\bibitem{KRY1963}A. K. Kerman, L. S. Rodberg  and J. E. Young, Phys. Rev. Lett. 11, 422 (1963).
\bibitem{Feshbach1993} H. Feshbach, Theoretical Nuclear Physics: Nuclear Reactions. (1993).
\bibitem{CH2013} L. F. Canto, and M. S. Hussein, ``Scattering Theory of Molecules, Atoms, and Nuclei", World Scientific, (2013).
\bibitem{Potel2015}G. Potel, F. M. Nunes, and I. J. Thompson, Phys. Rev. C 92, 034611 (2015).
\bibitem{Escher2012}J. E. Escher, J. T. Burke, F. S. Dietrich, N. D. Scielzo, I. J. Thompson, and W. Younes, Rev. Mod. Phys. 84, 353 (2012).
\bibitem{Carlson2015}B. V. Carlson, R. Capote, and M. Sin, arXiv:1508.01466 (2015).
\bibitem{Potel2016} G. Potel, G. Perdikakis, B. V. Carlson, J. Lei, A. M. Moro, J. E. Escher, W. Dickhoff, J. Rotureau, P. Capel, A. Macchiavelli, S. Pain, F. M. Nunes, and M. S. Hussein,
in preperation.
\bibitem{Serber1950} R. Serber, Phys. Rev. 80, 1098 (1950); Nature 166, 709 (1950); Proc. Roy. Soc. 208 A, 559 (1951).
\bibitem{Baur1978} A. Budzanowski, G. Baur, C. Alderliesten, J. Bojowald, C. Mayer-B\"oricke, W. Oelert, P. Turek, F. R\"{o}sel, and D. Trautmann, Phys. Rev. Lett. 41, 635 (1978).
\bibitem{Baur1980-1}G. Baur, R. Shyam, F. R\"{o}sel, and D. Trautmann, Phys. Rev. C 21, 2668 (1980).
\bibitem{Baur1980-2}R. Shyam, G. Baur, F. R\"{o}sel, and D. Trautmann, Phys. Rev. C 22, 1401 (1980).
\bibitem{BG06}C.A. Bertulani and A. Gade, Comp. Phys. Comm. 175, 372 (2006).
\bibitem{But50} S.T. Butler, Phys. Rev. 80, 1095 (1950); Nature 166, 709 (1950); Proc. Roy. Soc. 208 A, 559 (1951).
\end{thebibliography}
\end{document}